\documentclass[a4paper]{revtex4-1}
\usepackage{graphicx}
\usepackage{epstopdf,gensymb}
\usepackage{color}
\usepackage{subfigure}
\usepackage{bm}
\usepackage{float}

\begin{document}

\title{Novel Aspects of Direct Laser Acceleration of Relativistic Electrons}

\author{A. V. Arefiev$^1$, A. P. L. Robinson$^2$, and V. N. Khudik$^1$}
\affiliation{$^1$ Institute for Fusion Studies, The University of Texas, Austin,Texas 78712, USA\\$^2$ Central Laser Facility, STFC Rutherford-Appleton Laboratory, Didcot, OX11 0QX, UK}
\date{\today}

\begin{abstract}
We examine the impact of several factors on electron acceleration by a laser pulse and the resulting electron energy gain. Specifically, we consider the role played by: 1) static longitudinal electric field; 2) static transverse electric field; 3) electron injection into the laser pulse; and 4) static longitudinal magnetic field. It is shown that all of these factors lead, under certain conditions, to a considerable electron energy gain from the laser pulse. In contrast with other mechanisms such as wakefield acceleration, the static electric fields in this case do not directly transfer substantial energy to the electron. Instead, they reduce the longitudinal dephasing between the electron and the laser beam, which then allows the electron to gain extra energy from the beam. The mechanisms discussed here are relevant to experiments with under-dense gas jets, as well as to experiments with solid-density targets involving an extended pre-plasma. 
\end{abstract}

\maketitle

\section{Introduction}

The generation of super-thermal electrons is a commonly used approach in high intensity laser physics for coupling the incident laser power and then utilizing it for secondary particle and radiation sources. Examples of successfully employing this approach include: generation of energetic ions by solid-density foils~\citep{Flippo2010}, x-ray generation in gas jets~\citep{Kneip2008}, positron production in high-Z materials~\citep{Chen2010}, and neutron production from a metal converter~\citep{Pomerantz2014}. Laser interaction with electrons in the target strongly depends on the density of the target and the laser pulse duration. The interaction at sub-critical electron densities is volumetric and is not limited to the surface of the target. In this case, a propagating laser beam pulses electrons forward and radially outwards, causing cavitation of electron density. If the duration of the laser pulse is much shorter than the characteristic electron response time, then the laser pulse can generate a plasma structure that is moving together with the pulse. This is the so-called wakefield acceleration regime, where a forward-moving longitudinal field created in the plasma accelerates a bunch of electrons. If the duration of the laser pulse is much longer than the characteristic electron response time, then the laser pulse creates a quasi-static channel in the electron density that slowly evolves on an ion time-scale~\citep{Willingale2013}. The latter is the regime preferred for those applications that require copious energetic electrons, such as ion acceleration in the target-normal-sheath-acceleration regime~\citep{Flippo2010} and positron production~\citep{Chen2010}. It is important to point out that an extended under-dense plasma layer (preplasma) can be naturally generated at the front surface of the target as a result of a prepulse. Therefore, the interaction of a long laser pulse with an under-dense target is relevant not only to those experiments that utilize a gas jet~\citep{Mangles2005,Kneip2008,Willingale2013}, but also to the experiments with solid density targets~\citep{Flippo2010,Chen2010}.

A channel in the electron density that is created and sustained by a long laser beam has quasi-static transverse and longitudinal electric fields. The charge separation is balanced in this case by a transverse gradient of the ponderomotive pressure of the laser beam. An electron injected into the channel would be accelerated in the forward direction by the fields of the laser beam. However, this acceleration takes place in the presence of the channel's quasi-static electric field. An oft-cited acceleration process in this configuration is the `betatron resonance' acceleration process \citep{Pukhov1999}, although the complex and subtle physics of this configuration means that further examination is required \citep{Arefiev2012}. In this paper, we examine single electron dynamics in such a channel and determine the conditions for significant enhancement in electron energy gain as compared to the case of electron acceleration in a vacuum. Other scenarios of electron acceleration in sub-critical plasmas have also been proposed and researched \citep{Meyer-ter-Vehn1999,Sheng2002,Chen2006,Gallard2011,Paradkar2012,Naseri2012,Willingale2013,Krygier2014}. The goal of this manuscript is not to provide a comprehensive overview of these scenarios, but rather to present a structured review of our recent research and new findings regarding electron acceleration in a steady-state plasma channel.

We begin by reviewing our recent findings~\citep{Arefiev2012,Robinson2013,Arefiev2014} that show that quasi-static transverse and longitudinal electric fields that naturally arise in the plasma can lead to a significant electron energy gain, with the maximum electron energy exceeding the ponderomotive energy. We then examine how electron injection into the laser beam affects the electron energy gain by considering electrons that begin their motion at different phases of the wave. It is shown that a favorable initial phase can allow continuous electron energy gain over multiple wave periods, with the peak energy greatly exceeding what is expected for an electron that is initially at rest and irradiated by a wave packet with a gradually increasing amplitude~\citep{Arefiev2014}. We also examine the role played by a longitudinal externally applied magnetic field. It is shown that if an electron is either pre-accelerated or is accelerated by a longitudinal electric field then this can drive it into resonance, and electron cyclotron absorption can then occur.

The rest of the paper is organized as follows. Section~\ref{Sec2} presents a particle-in-cell simulation for a two-dimensional set-up that shows the key features of the channel formed in the plasma by the laser. This establishes the context for the analysis that follows. In Section~\ref{Sec3} we then formulate a single electron model that can be used to examine the role of different effects on electron dynamics in a quasi-static channel. In the following sections, we examine the roles played by: longitudinal (Sec.~\ref{Sec4}) and transverse (Sec.~\ref{Sec5}) static electric fields; initial electron injection (Sec.~\ref{Sec6}); static longitudinal magnetic field (Sec.~\ref{Sec7}). Finally, in Sec. (Sec.~\ref{Sec8}) we summarize our findings and discuss several issues that need to be addressed in the future regarding electron acceleration in long plasma channels. Sections \ref{Sec3} - \ref{Sec5} are essentially providing a review and summary of our recent work, whereas in Sections \ref{Sec2}, \ref{Sec6}, and \ref{Sec7} we are presenting new results.

%+++++++++++++++++++++++++++++++++++++++++++++++++++++++++

\begin{figure}
  \subfigure
  {\includegraphics[width=7cm]{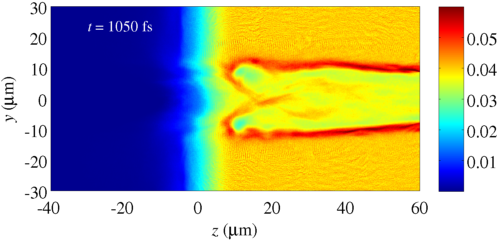}} 
  \quad
  \subfigure
  {\includegraphics[width=7cm]{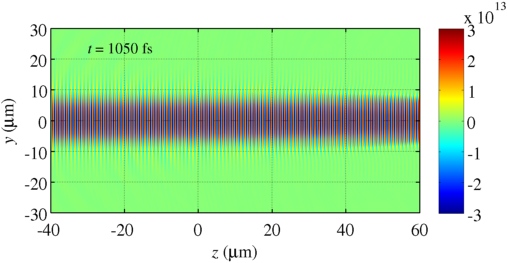}}
\label{fig:1} \caption{(Left) Electron density and (Right) laser electric field after a steady-state channel is established in the plasma. The electron density is averaged over ten laser periods and normalized to the critical density, $n_{crit} = 11.1 \times 10^{26}$ m$^{-3}$. The plot of the laser electric field is a snapshot of the $x$-component, given in volt/m.}
\end{figure}

\section{Channel formation in an under-dense plasma} \label{Sec2}

In order to illustrate the key features of the channel formed in the plasma, we have carried out a 2D-3V particle-in-cell simulation where a laser beam is normally incident onto an extended significantly under-dense plasma with cold electrons. The ions are deliberately treated as immobile in this simulation to prevent gradual channel expansion and eliminate effects associated with ion mobility. The length and width of the domain is 200 $\mu$m $\times$ 160 $\mu$m (12000 $\times$ 1600 cells). The longitudinal cell size satisfies the criterion formulated by \citet{Arefiev2015}, which ensures that the electron acceleration by the laser pulse is simulated correctly. The laser pulse propagates along the $z$-axis and its focal plane is located at $z = 0$ $\mu$m. The maximum laser amplitude in the focal plane rises over 150 fs to $a_0 = 10$ and then remains constant. The laser wave-length is $\lambda = 1$ $\mu$m. The FWHM for the laser intensity in the focal plane is 10 $\mu$m. The laser electric field is polarized out of the plane of the simulation (directed along the $x$-axis). The maximum plasma (electron) density is $n_0 = 6.7 \times 10^{25}$ m$^{-3}$, which is 6\% of the critical density for the considered wave-length. The plasma density rises gradually from 0 to $n_0$ over 30 $\mu$m [$-20$ $\mu$m $<z<10$ $\mu$m].  The plasma extends longitudinally to $z = 130$ $\mu$m. In the transverse direction, the plasma width is 100 $\mu$m, with $|y| \leq 50$ $\mu$m. We use 10 macro-particles per cell to represent electrons and 5 macro-particles per cell to represent ions. 

The incident laser pulse is able to penetrate the plasma, because the electron density is considerably under-dense. The gradient of the ponderomotive pressure at the leading edge causes the electrons to move forward and outwards. The uncompensated ion charge provides a counteracting force, which causes significant fluctuations of the electron density. After the initial transition that lasts multiple plasma periods, the plasma eventually reaches a steady-state equilibrium shown in Figure \ref{fig:1}.

\begin{figure}
  \subfigure
  {\includegraphics[width=7cm]{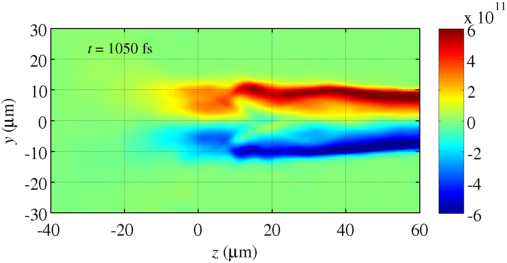}} 
  \quad
  \subfigure
  {\includegraphics[width=7cm]{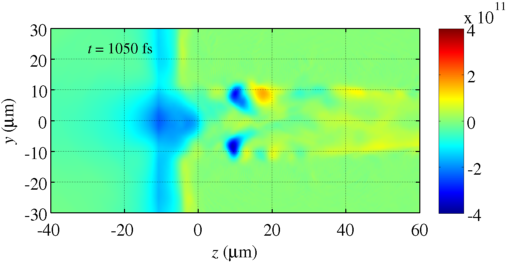}}
\label{fig:2} \caption{(Left) Time-averaged transverse and (Right) longitudinal components of the electric field in the steady-state channel. The fields have been averaged over ten laser periods. The units are volt/m.}
\end{figure}

In the equilibrium, the laser pulse maintains a channel in the electron density. The channel has a coaxial structure with a positively charged core where the electron density is depleted and a negatively charged outer shell where there is excess electron density and where the electron density peaks. There is a resulting steady-state transverse electric field (see Fig. \ref{fig:2}) that is directed outwards from the axis of the channel ($y = 0$ $\mu$m). The charge of the outer shell fully compensates the positive charge of the core, so that the transverse electric field vanishes outside of the shell. 

There is also a considerable negative electric field at the channel entrance, as seen in Fig. \ref{fig:2}. This is essentially a boundary effect resulting from the coaxial structure of the channel with a positively charged core and a negatively charged shell. The radius of the channel changes in the longitudinal direction on a scale much greater than the radius itself, which causes the longitudinal electric field to vanish inside the channel away from the entrance. This field structure shown in Fig. \ref{fig:2} changes insignificantly with time after the channel has been established. 

\begin{figure}
  \subfigure
  {\includegraphics[width=7cm]{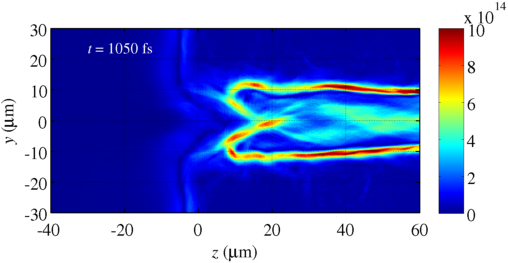}} 
  \quad
  \subfigure
  {\includegraphics[width=7cm]{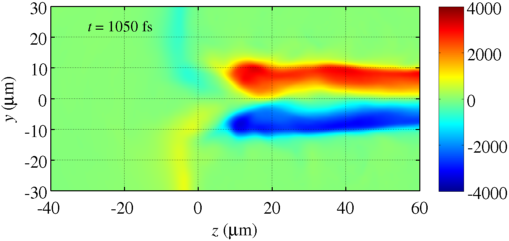}}
\caption{(Left) Time-averaged plasma current density and (Right) magnetic field in the steady-state channel. The averaging was performed over ten laser periods. The plotted component of the magnetic field is the $x$-component, given in Tesla. The current density is given in A/m$^2$.} \label{fig:3}
\end{figure}

Even though the electron density shown in Fig. \ref{fig:1} is in a steady-state, there is a net longitudinal motion of electrons inside the channel in the direction of the laser pulse propagation. Figure~\ref{fig:3} shows that the laser drives a negative longitudinal current in the core of the channel. This is a current of energetic electrons accelerated by the laser. This current generates a steady-state magnetic field in the channel (see Fig.~\ref{fig:3}), directed along the $z$-axis (out of the plane of the simulation). There is also a return current that runs in the outer channel shell and that allows for the plasma configuration to remain in a steady-state. The current loop closes at the channel entrance. 

In terms of generating highly energetic electrons, the following is observed. New electrons are continuously injected into the channel at the channel opening, as evident from the current configuration in Fig.~\ref{fig:3}. The injected electrons are subsequently accelerated by the laser pulse in the forward direction. The magnetic field that is produced by the current of these electrons effectively prevents injection of new electrons into the channel away from the channel opening. The most highly energetic electrons observed in this system are accelerated in, and along the length of, the channel. The interpretation of these simulations therefore requires one to build a framework that explains the acceleration along the length of the channel.

To conclude, a sufficiently long laser pulse produces a steady-state channel in an under-dense plasma with static longitudinal and transverse electric fields. The laser pulse accelerates electrons in the channel in the forward direction. The electron population in the channel is replenished, as new electrons are continuously injected into the channel near its opening. The presented simulation illustrates general features of an interaction between a plasma and a relatively long laser pulse. Understanding energetic electron generation in such setup requires that one understands the electron acceleration mechanisms in the channel.

%+++++++++++++++++++++++++++++++++++++++++++++++++++++++++

\section{Single electron model} \label{Sec3}

In order to examine the role of static electric and magnetic fields that can be present in a channel on electron dynamics, it is helpful to consider a simplified setup where a single electron is irradiated by a linearly polarized plane electromagnetic wave in a fully evacuated steady-state ion channel. The main advantage of such a single electron model is that the problem reduces to finding the electron dynamics in {\it{given}} electric, $\bf{E}$, and magnetic, $\bf{B}$ fields. These fields can be a superposition of the fields of the channel, the fields of the wave, and an externally applied magnetic field. The equations that describe the electron dynamics are
\begin{eqnarray}
&& \frac{d {\bf{r}}}{d t} = \frac{c}{\gamma} \frac{{\bf{p}}}{m_e c}, \label{main_eq:1}\\
&& \frac{d {\bf{p}}}{d t} = - |e| {\bf{E}} - \frac{|e|}{\gamma m_e c} [{\bf{p}} \times {\bf{B}}],  \label{main_eq:2}
\end{eqnarray}
where ${\bf{r}}$ and ${\bf{p}}$ are the electron position and momentum, $t$ is the time in the channel frame of reference, $e$ and $m_e$ are the electron charge and mass, $c$ is the speed of light, and 
\begin{equation}
\gamma = \sqrt{1+ p^2 /m_e^2 c^2}
\end{equation}
is the relativistic $\gamma$-factor.

The objective of this manuscript is to elucidate various factors that can impact electron dynamics in a plasma channel rather than to develop a comprehensive treatment incorporating all of the relevant physics. With this in mind, we proceed by formulating reduced and easily tractable sets of equations in the following sections for various cases. In all of the cases, we consider a single electron irradiated by a plane electromagnetic wave propagating along the $z$-axis. Similar to~\citet{Arefiev2012,Arefiev2014}, we assume a slab-like (2D Cartesian) ion channel. The $(x,z)$-plane at $y=0$ is the midplane of the ion slab, with the $y$-axis directed across the slab. The direction of the laser electric field is specified by a polarization angle $\theta$. 

The laser wave field is described by a normalized vector potential
\begin{equation} \label{eq:a}
{\bf{a}} (z,t) = a(\xi) \left[ {\bf{e}}_x \cos \theta   + {\bf{e}}_y \sin \theta \right]
\end{equation}
that is only a function of a dimensionless phase variable
\begin{equation} \label{Eq_xi}
	\xi \equiv \omega (t - z / c) ,
\end{equation}
where $\omega$ is the wave frequency and ${\bf{e}}_x$ and ${\bf{e}}_y$ are unit vectors. In the analysis that follows, we consider pulses with
\begin{eqnarray} \label{pulse_shape}
&& a(\xi) = a_*(\xi) \sin(\xi), \\
&& 0 \leq a_* \leq a_0,
\end{eqnarray}
where $a_*(\xi)$ is a given slowly varying envelope. The electric and magnetic fields of the laser pulse are then given by
\begin{eqnarray}
&& {\bf{E}}_{wave} = - \frac{m_e c}{|e|} \frac{\partial {\bf{a}}}{\partial t} = - \frac{m_e \omega c}{|e|} \frac{d {\bf{a}}}{d \xi}, \label{wave:1} \\
&& {\bf{B}}_{wave} = \frac{m_e c^2}{|e|}  [\nabla \times {\bf{a}}] = [{\bf{e}}_z \times {\bf{E}}_{wave}], \label{wave:2}
\end{eqnarray}
where ${\bf{e}}_z$ is a unit vector. The phase of the laser field changes in time at the rate
\begin{equation} \label{eq:deph}
\frac{d \xi}{d t} = \frac{\omega}{\gamma} \left( \gamma - \frac{p_z}{m_e c} \right).
\end{equation}
This equation must be solved together with Eqs.~(\ref{main_eq:1}) and (\ref{main_eq:2}) in order to describe the time evolution of the laser fields at the electron location.

A very similar approach is taken in the textbook treatment of \citet{Boyd2003}, which only deals with the case of an initially immobile electron interacting with a plane electromagnetic wave alone. This case is still worth discussing briefly to provide context for the discussion that follows.  For the case of the polarization angle $\theta =0$, one obtains the following solution for the electron momentum,
\begin{eqnarray}
&& \frac{p_x}{m_ec} = a, \\
&& \frac{p_z}{m_ec} = \frac{a^2}{2}.
\end{eqnarray}
Importantly, this solution is obtained by first finding two constants of motion for this problem, namely,
\begin{eqnarray}
\label{eq:com1}
&& \frac{p_x}{m_ec} - a = 0, \label{eq:com2_1} \\
&& \gamma - \frac{p_z}{m_ec} = 1. \label{eq:com2}
\end{eqnarray}
Equation (\ref{eq:com1}) is a consequence of conservation of canonical momentum, and has a direct bearing on the transverse momenta that are achieved. Equation (\ref{eq:com2}) should be understood, via comparison to Eq.~(\ref{eq:deph}), to be a dephasing rate. We see that the plane wave case puts strict limits on the maximum momenta that can be achieved. It is also clear that there is no retention of momentum after the electron leaves the laser field in agreement with the Lawson-Woodward theorem. %Clearly, this interaction cannot be termed DLA as the electron ultimately do not gain any energy.

In what follows, we will add imposed electric and magnetic fields to this problem that represent the fields that are self-consistently generated in real laser-plasma interactions.  We then look to see how the introduction of these fields changes the two crucial constants of motion given by Eqs.~(\ref{eq:com2_1}) and (\ref{eq:com2}), how this affects the maximum momenta that can be achieved, and how this affects the adiabaticity of the system. The introduction of these additional fields alters the problem, and opens up the possibility of the electrons acquiring net energy from the interaction. A discussion of other single particle approaches can be found in \citep{Pukhov2003,Tsakiris2000}.

%A very elementary reason for this is that the electron in the channel fields is an oscillator, and coupling two oscillators allows, in general, the net transfer of energy.  Such a system can now be termed DLA. A discussion of this and other single particle approaches can be found in \cite{Pukhov2003,Tsakiris2000}.

%+++++++++++++++++++++++++++++++++++++++++++++++++++++++++

\section{Role of a longitudinal static electric field} \label{Sec4}

We start with a case where, in addition to the laser field, there is only a static longitudinal electric field $E_0$ present in some region along the $z$-axis. The content of this section is based on the findings of \citet{Robinson2013}. Without loss of generality, we assume that the laser electric field is polarized along the $x$-axis [which corresponds to $\theta = 0$ in (\ref{eq:a})]. It follows from Eq.~(\ref{main_eq:2}) that
\begin{eqnarray}
&& \frac{d}{d t} \left( \frac{p_x}{m_e c} - a \right) = 0, \label{Eq_long:1} \\
&& \frac{d}{d t} \left( \frac{p_y}{m_e c} \right) = 0, \label{Eq_long:2} \\
&& \frac{d}{d t} \left( \frac{p_z}{m_e c} \right)  = - \frac{|e| E_0}{m_e c} + \frac{\omega}{\gamma}  \frac{p_x}{m_e c} \frac{da}{d \xi}, \label{Eq_long:3}
\end{eqnarray}
where we took into account the expressions for the wave electric and magnetic fields given by Eqs.~(\ref{wave:1}) and (\ref{wave:2}). 

It can be directly shown using Eq.~(\ref{eq:deph}) and Eqs.~(\ref{Eq_long:1}) - (\ref{Eq_long:3}) that 
\begin{equation} \label{eq:deph:1}
\frac{d}{dt} \left( \gamma - \frac{p_z}{m_e c}\right) = \frac{|e| E_0}{\omega m_ e c} \frac{d \xi}{dt}.
\end{equation}
Introducing proper time $\tau$ defined by the relation
\begin{equation} \label{eq:4.5}
d\tau / dt = 1/\gamma,
\end{equation}
we can rewrite Eq.~(\ref{eq:deph}) as
\begin{equation} \label{eq:deph_2}
\frac{1}{\omega} \frac{d \xi}{d \tau} = \gamma - \frac{p_z}{m_e c}.
\end{equation}
Therefore, the quantity $\gamma - p_z/m_e c$ can be interpreted as a dephasing rate, as it gives the rate at which the phase of the wave field sampled by the electron changes in an instantaneous frame of reference where the electron is at rest. In what follows, we will use the notation
\begin{equation} \label{eq:4.7}
R \equiv \frac{1}{\omega} \frac{d \xi}{d \tau}
\end{equation}
and refer to $R$ as a dephasing rate.

In the absence of a longitudinal electric field, the dephasing rate $R$ remains constant, since in this case $\gamma - p_z/m_e c$ is an integral of motion according to Eq.~(\ref{eq:deph:1}). The role of the longitudinal electric field is then to either decrease or increase the dephasing rate depending on its direction. Note that $d\xi/dt$ is always positive because the electron moves with velocity slower than $c$. Taking into account that the laser pulse propagates in the positive direction along the $z$-axis, we conclude that a longitudinal electric field pointing against the direction of the pulse propagation decreases the dephasing rate, whereas a longitudinal electric field pointing in the direction of the pulse propagation increases the dephasing rate.

Our next step is to determine how a change in the dephasing impacts the electron acceleration and the resulting energy gain from the laser field. It is instructive to first examine the electron dynamics in the absence of a static longitudinal electric field. We assume that an electron is irradiated by a laser pulse with a gradually increasing envelope $a_*(\xi)$. Without any loss of generality, we assume that the electron is located at $z=0$ at $t=0$, which corresponds to $\xi=0$. The laser pulse arrives at $\xi > 0$, with $a(\xi) = 0$ for $\xi \leq 0$. The dephasing rate in this case is determined exclusively by the initial $(t=0)$ electron momentum, ${\bf{p}} (t=0) \equiv {\bf{g}}$:
\begin{equation} \label{eq:4.8}
R = \sqrt{1+g^2} - g_z.
\end{equation}
It is straightforward to integrate Eqs.~(\ref{Eq_long:1}) and (\ref{Eq_long:2}) and combine the resulting expressions with the relation $\gamma - p_z/m_e c = R$ to find that
\begin{eqnarray}
&& p_x = g_x + am_ec, \label{eq:4.9}\\
&& p_y = g_y, \\
&& p_z = \left[ 1+ \left(\frac{g_x}{m_ec} + a \right)^2+ \left(\frac{g_y}{m_ec} \right)^2 - R^2 \right] \frac{m_ec}{2R}, \label{eq:4.11} \\
&& \gamma = \left[ 1+ \left(\frac{g_x}{m_ec} + a \right)^2+ \left(\frac{g_y}{m_ec} \right)^2 + R^2 \right] \frac{1}{2R}. \label{eq:4.12}
\end{eqnarray}

\begin{figure}
  \subfigure
  {\includegraphics[width=6cm]{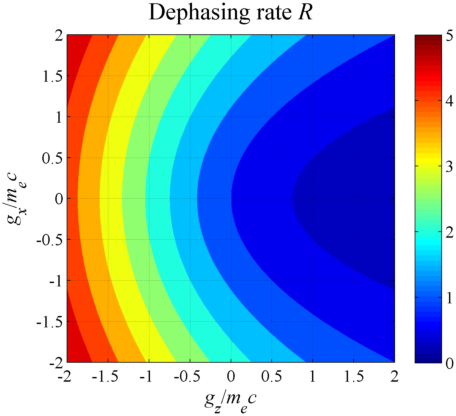}} 
  \quad
  \subfigure
  {\includegraphics[width=6cm]{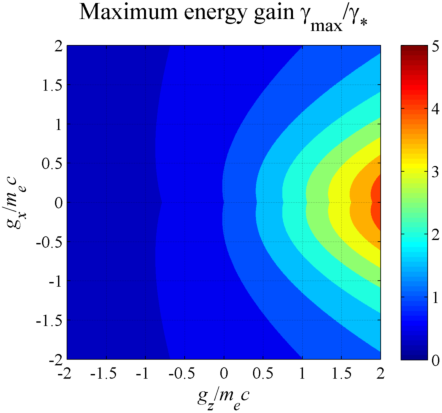}}
\label{fig:vac} \caption{(Left) Electron dephasing rate $R$ and (Right) the corresponding maximum $\gamma$-factor $\gamma_{\max}$ achieved during acceleration in the laser pulse for different values of $x$ and $z$ components of the electron momentum ($g_x$ and $g_z$). Here $\gamma_{\max}$ is calculated for $a_0 = 10$ and it is normalized to $\gamma_* = 1+a_0^2/2$.}
\end{figure}

An electron that is initially at rest can achieve the maximum of 
\begin{equation}
\gamma = \gamma_* \equiv 1 + a_0^2/2
\end{equation}
accelerating in the laser field of amplitude $a_0$. The electron motion is strongly relativistic for waves with $a_0 \gg 1$. Most of the electron energy at these wave amplitudes is associated with the longitudinal motion, since $\max p_z / \max p_x = a_0/2 \gg 1$. One might expect that a change in the initial momentum would have little impact on the maximum $\gamma$-factor as long as the initial $\gamma$ is much less than $\gamma_*$. To show that this is not the case, let us consider a wave with amplitude $a_0 \gg 1$ that irradiates an electron with $|g_x/m_ec| \ll a_0$, $|g_y/m_ec| \ll a_0$, and $|g_z/m_ec| \ll a_0^2/2$. In this case, the maximum $\gamma$-factor according to Eq.~(\ref{eq:4.12}) is approximately $\gamma \approx a_0^2/2R$. If the initial electron momentum is non-relativistic, then it follows from Eq.~(\ref{eq:4.8}) that $R \approx 1$ and the maximum $\gamma$-factor would be equal to that of an electron that is initially at rest ($\gamma_*$). However, if the initial electron momentum is relativistic, then the dephasing rate can change considerably. For a given initial momentum $q \gg m_e c$, the lowest value of the dephasing is $R \approx 1/2g$ for $g_z = g$ and the highest value of the dephasing is $R \approx 2g$ for $g_z = -g$. Therefore, the maximum $\gamma$-factor that the electron can achieve in a laser pulse with amplitude $a_0 \gg 1$ changes depending on the orientation of the initial electron momentum in the range
\begin{equation}
\frac{a_0^2}{4g} \leq \gamma_{\max} \leq g a_0^2,
\end{equation}
where it is assumed that $g \gg 1$. This aspect is illustrated in Fig.~\ref{fig:vac}, where the panel on the left shows the dephasing rate $R$ as a function of the initial electron momentum and the panel on the right shows the corresponding $\gamma_{\max}/\gamma_*$ for $a_0 = 10$. The value of $\gamma_{\max}$ is calculated using Eq.~(\ref{eq:4.12}).

This result can be qualitatively understood by considering the longitudinal momentum balance equation (\ref{Eq_long:3}) for an electron that is initially moving along the $z$-axis. The transverse momentum for this electron is $p_x = a m_ec$ and Eq.~(\ref{Eq_long:3}) can thus be rewritten as
\begin{equation} \label{eq:4.15}
\frac{d}{d \tau} \left( \frac{p_z}{m_e c} \right)  =\omega  \frac{d}{d \xi} \left( \frac{a^2}{2} \right),
\end{equation}
where we also took into account the relation~(\ref{eq:4.5}) to replace $t$ with $\tau$. Equation (\ref{eq:4.15}) shows that the longitudinal acceleration of the electron is caused by a gradient of the wave amplitude. The corresponding longitudinal force oscillates as $\sin(2\xi)$, changing its sign every time $\xi$ increases by $\pi$.
Therefore, the longitudinal acceleration is limited by how fast $\xi$ changes with $\tau$. By definition, this rate of change is the dephasing rate $R$. In essence, if $R$ is decreased, the electron can stay in phase with the wave longer and can gain more energy before being decelerated.

We can now use the already obtained results to quantify the effect of a longitudinal electric field $E_0$ on electron dynamics. Let us consider a case where the applied field $E_0$ is constant over a region of length $l$. It then follows from Eq.~(\ref{eq:deph:1}) that in this region
\begin{equation} \label{eq:4.16}
\frac{d}{dt} \left( \gamma - \frac{p_z}{m_e c} - \frac{|e| E_0 \xi}{\omega m_ e c} \right) = 0.
\end{equation}
We integrate this equation and use the definition $R \equiv \gamma - p_z/m_ec$ to find that 
\begin{equation} \label{eq:4.17}
\Delta R = \frac{|e| E_0}{\omega m_ e c} \Delta \xi,
\end{equation}
where $\Delta \xi$ is the change in wave phase that accumulates as the electron traverses the region with the electric field. If the change in the dephasing rate $\Delta R$ caused by the electric field $E_0$ is considerable compared to the dephasing rate $R$ prior to the interaction with the electric field, then the subsequent electron energy gain will change considerably as well. This result directly follows from Eqs.~(\ref{eq:4.11}) and (\ref{eq:4.12}), where $R$ should be replaced with $R+\Delta R$.

\begin{figure}
  \subfigure
  {\includegraphics[width=6.5cm]{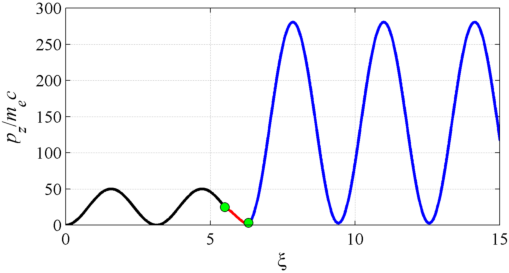}} 
  \quad
  \subfigure
  {\includegraphics[width=6.5cm]{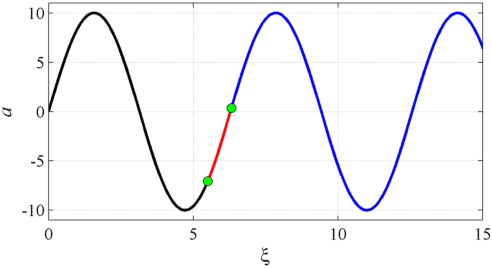}}

  \subfigure
  {\includegraphics[width=6.5cm]{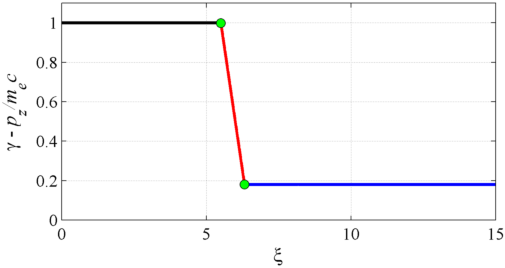}} 
  \quad
  \subfigure
  {\includegraphics[width=6.5cm]{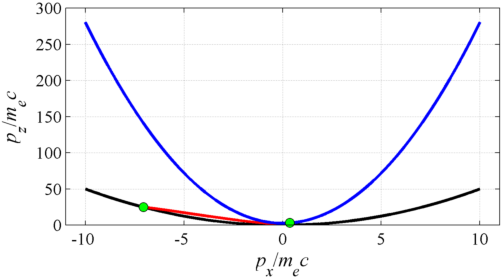}}
\caption{Dynamics of an initially immobile electron irradiated by a plane wave with $a_0=10$ that passes through a region with a longitudinal electric field $|e| E_0 /m_e \omega c = - 0.1 a_0$ located at $23.87 \leq z/\lambda \leq 26.27$. The dynamics before, during, and after the interaction with the field is shown in black, red, and blue.} \label{fig:5}
\end{figure}

An example of such an interaction with a longitudinal electric field is shown in Fig.~\ref{fig:5}. In this case, a wave of amplitude $a_0=10$ irradiates an electron that was initially at rest. Prior to the interaction with the longitudinal field, the electron moves along a parabola in the momentum space (black curve in lower-right panel) with $p_z/m_ec \leq a_0^2/2$ and $|p_x/m_ec| \leq a_0$. The dephasing rate is constant and equal to unity, $R = \gamma - p_z/m_ec = 1$ (black line in upper-right panel). The electron encounters a longitudinal electric field $|e| E_0 /m_e \omega c = - 0.1 a_0$ at $z = 23.87 \lambda$ in a region of width $l = 2.4 \lambda$. This longitudinal electric field has an amplitude that is ten times smaller than the maximum laser electric field. The evolution of all the quantities during the interaction with the longitudinal field is shown by red curves in all four panels. The dephasing drops considerably from $R = 1$ to $R = 0.18$. At the same time, the longitudinal and transverse components of the momentum decrease as well. Electron trajectory in the momentum space following the interaction with the longitudinal field is shown with a blue curve in the lower-right panel of Fig.~\ref{fig:5}. The electron again moves along a parabola, but the new parabola is steeper than the one prior to the interaction with the field. The maximum longitudinal momentum is enhanced by roughly a factor of six. 

We can then conclude that the role of the longitudinal electric field is to launch the electron onto an energetic trajectory. This is achieved by decreasing the dephasing rate rather than by transferring a considerable amount of energy to the electron during the interaction. The considerable energy gain takes place {\it{after}} the interaction with the longitudinal field, with the extra energy being transferred to the electron from the laser pulse and not from the longitudinal field.

\begin{figure}
  \subfigure
  {\includegraphics[width=6.5cm]{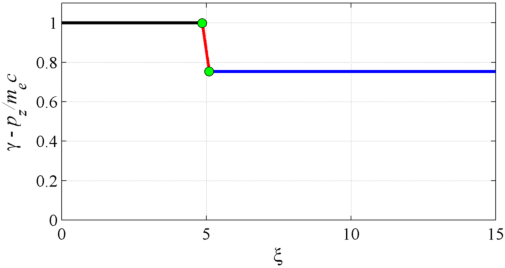}} 
  \quad
  \subfigure
  {\includegraphics[width=6.5cm]{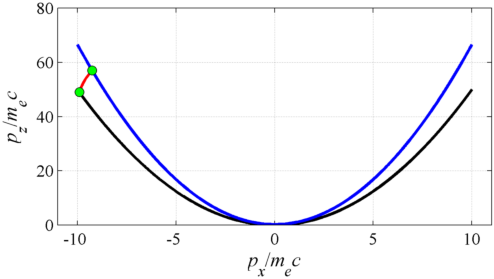}}
\caption{Electron dynamics for the same set-up as in Fig.~\ref{fig:5} with the only difference that the region with the longitudinal field has been shifted by $\Delta z = -3.98 \lambda$ to $19.89 \leq z/\lambda \leq 22.29$. The dynamics before, during, and after the interaction with the field is shown in black, red, and blue.} \label{fig:6}
\end{figure}

The timing of the interaction with the longitudinal field has a strong effect on the dephasing rate and, as a result, on the subsequent energy gain from the laser pulse. This can be illustrated by comparing Figs.~\ref{fig:5} and \ref{fig:6}. The only parameter that was changed to generate Fig.~\ref{fig:6} is the location of the region with the longitudinal field, while the field magnitude and the length of the region are kept constant. The region now begins at $z = 19.89 \lambda$ as opposed to $z = 23.87 \lambda$. In Figure~\ref{fig:6}, the dephasing is reduced only to $R = 0.75$ and, as a result, the maximum longitudinal momentum after the interaction increases only to $66.6 m_ec$ (in contrast to $280.5 m_ec$ in Fig.~\ref{fig:5}). 

In order to explain why the timing of the interaction is important, let us first consider the electron motion across the same region of length $l$ but in the absence of the longitudinal field. As the electron moves across the region, the wave phase increases with time according to Eq.~(\ref{eq:deph}). The total travel time is $\delta t \approx l/c$ 
for a relativistically moving electron. We set $\gamma - p_z/m_ec = 1$ and assume for simplicity that the $\gamma$-factor does not change significantly to find from Eq.~(\ref{eq:deph}) that the wave phase increases by 
\begin{equation}
\Delta \xi \approx \omega c/\gamma l.
\end{equation}
The key conclusion is that $\Delta \xi$ decreases with the increase of the electron $\gamma$-factor. The same trend holds as we compare Figs.~\ref{fig:5} and \ref{fig:6}. Changing in the location of the region with the electric field we effectively change the $\gamma$-factor and $p_z$ of the electron as it approaches the region. The $\gamma$-factor is considerably higher in the case presented in Fig.~\ref{fig:6} and, as a result, $\Delta \xi$ decreases, as evident from comparing the width of the red segments in the plots of $\gamma - p_z/m_e c$ in Figs.~\ref{fig:5} and \ref{fig:6}. On the other hand, the electric field is the same in both cases and that is why the decrease in the dephasing rate $R$ given by Eq.~(\ref{eq:4.17}) is less for the regime of Fig.~\ref{fig:6}. Therefore, higher $\gamma$-factor {\it{prior}} to the interaction with the longitudinal field leads to a reduced electron gain during the acceleration by the wave {\it{after}} the interaction.

In conclusion, we have shown that a relatively weak static longitudinal electric field can have a strong effect on the dynamics of an electron accelerating in a laser pulse. The field affects the electron dynamics by changing the dephasing rate. A field directed against the laser pulse propagation can significantly decrease the dephasing rate without having to transfer a considerable amount of energy to the electron during the interaction. However, a considerable energy gain takes place {\it{after}} the interaction with the longitudinal field, with the extra energy being transferred to the electron from the laser pulse.

%+++++++++++++++++++++++++++++++++++++++++++++++++++++++++

\section{Role of a transverse static electric field} \label{Sec5}

We now consider a case where an electron is irradiated by a plane wave in an axially uniform ion channel, so that there is only a transverse electric field in addition to the laser field. The content of this section is based on the findings of \citet{Arefiev2012,Arefiev2014}. In the 2D channel described in Sec.~\ref{Sec3}, the electric field has only a $y$-component $E = 4 \pi n_0 |e| y$, where $n_0$ is the density of the singly charged ions in the channel. It follows from Eqs.~(\ref{main_eq:1}) and (\ref{main_eq:2}) that the equations that govern electron motion in the $(y,z)$-plane of such a channel are
\begin{eqnarray}
&& \frac{1}{\omega} \frac{d }{d \tau} \left( \frac{p_y}{m_e c} - a \sin \theta \right) = - \gamma \frac{\omega_p^2}{\omega^2}  \frac{\omega}{c} y, \label{Eq1} \\
&& \frac{1}{\omega} \frac{d}{d \tau} \left( \frac{p_z}{m_e c} \right)  = \left( \frac{p_y}{m_e c} - a \sin \theta \right) \sin \theta \frac{d a}{d \xi} + \frac{d}{d \xi} \left( \frac{a^2}{2} \right) , \label{Eq2}\\
&& \frac{1}{\omega} \frac{d}{d \tau} \left( \frac{\omega}{c} y \right) =  \frac{p_y}{m_e c} , \label{Eq3}\\
&& \frac{1}{\omega} \frac{d \xi}{d \tau} =  \gamma - \frac{p_z}{m_e c}, \label{Eq4}
\end{eqnarray}
where $p_y$ and $p_z$ are components of the electron momentum,
\begin{equation}
\gamma = \sqrt{1 + a^2 \cos^2 \theta + \left(p_y/m_e c\right)^2 + \left(p_z / m_e c \right)^2} \label{Eq5}
\end{equation}
is the relativistic $\gamma$-factor, and $\tau$ is proper time defined by the relation 
\begin{equation} \label{Eq6}
	d \tau /d t = 1 / \gamma.
\end{equation}
Here $\omega_p \equiv \sqrt{4 \pi n_0 e^2/m_e}$ is the plasma frequency. Equations~(\ref{Eq1}) and (\ref{Eq2}) are transverse and parallel momentum balance equations, whereas Eqs.~(\ref{Eq3}) and (\ref{Eq4}) relate the time evolution of the corresponding transverse and longitudinal coordinates. These equations assume that the electron is not moving along the $x$-axis before the laser pulse arrives and, as a result, $p_x - m_e c a \cos \theta = 0$ during the electron motion in the wave.

The transverse static electric field is axially uniform and, as a consequence of this, Eqs. (\ref{Eq1}) - (\ref{Eq4}) have the following integral of motion
\begin{equation} \label{R_main_0}
I \equiv \gamma - \frac{p_z}{m_e c} + \frac{\omega_p^2}{c^2} \frac{y^2}{2}.
\end{equation}
This integral of motion relates the dephasing rate $R = \gamma - p_z/m_ec$ that was introduced in Sec.~\ref{Sec4} to the amplitude of electron oscillations across the channel. The relation (\ref{R_main_0}) indicates that the dephasing rate can be dramatically decreased by amplifying the electron oscillations across the channel. If the amplitude of the transverse oscillations is amplified and approaches 
\begin{eqnarray} \label{y_crit}
&& y_* \equiv \left. \sqrt{2I} c \right/ \omega_p,
\end{eqnarray}
then $\gamma - p_z/m_e c$ becomes vanishingly small. As we have seen in Sec.~\ref{Sec4}, a decrease in the dephasing leads to an energy increase of the electron. Therefore, the relation (\ref{R_main_0}) suggests that amplification of transverse electron oscillations might be beneficial for increasing electron energy gain from the laser pulse.

\begin{figure}
  \subfigure
  {\includegraphics[width=6.5cm]{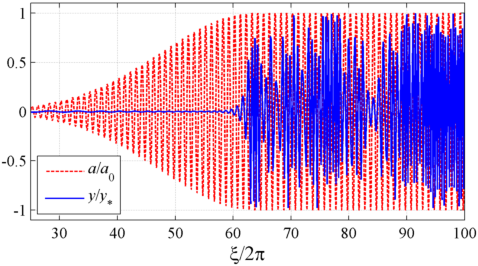}} 
  \quad
  \subfigure
  {\includegraphics[width=6.5cm]{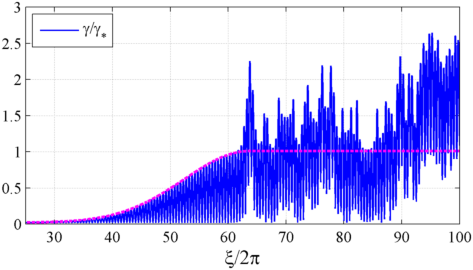}}
\caption{Electron dynamics in the regime where the oscillations are unstable ($a_0 = 10$ and $\omega_p/\omega = 0.175$) and there is no driving electric field across the channel. The dasher (red) curve in the left panel is the amplitude of the laser pulse. The dashed (magenta) curve in the right panel is the maximum $\gamma$-factor in the vacuum regime (without the static field of the channel).} \label{fig:100}
\end{figure}

A direct way to amplify the transverse oscillations is by applying a driving electric field across the channel. However, the transverse oscillations can be amplified even in the absence of a driving electric field. The amplification occurs in this case because electron oscillations across the channel are unstable in the presence of a linearly polarized wave. We proceed by first illustrating this aspect and then by examining how the amplification of transverse oscillations affects the electron energy gain.

We consider a single electron that performs small amplitude, $|y| \ll y_*$, non-relativistic oscillations across the channel before being irradiated by a laser pulse whose electric field is directed along the $x$-axis. As the laser arrives, it begins to drive electron oscillations along the $x$-axis, while pushing the electron forward. Note that there is no laser electric field across the channel. Since the oscillations across the channel are small, the electron motion in the $(x,z)$-plane is essentially unaffected by the channel and it resembles the vacuum case considered in Sec.~\ref{Sec4} where the electron moves only under the effect of the laser field. For high-amplitude waves with $a_0 \gg 1$, this motion is relativistic and it determines the electron $\gamma$-factor. Treating the motion in the $(x,z)$-plane as given, we can now examine its effect on the motion across the channel. 

The equation of motion across the channel follows directly from Eqs.~(\ref{Eq1}) and (\ref{Eq3}) by setting $\theta = 0$,
\begin{equation} \label{eq:5.9}
\frac{d^2 y}{d \tau^2} + \gamma \omega_p^2 y = 0.
\end{equation}
If the full electron motion is non-relativistic, then the motion across the channel is decoupled from the motion in the $(x,z)$-plane. In this case, the electron oscillates with frequency $\omega_p$ across the channel.
However, if the motion in the $(x,z)$-plane is relativistic, the motion across the channel is no longer decoupled. The 
latter is influenced by the former through the $\gamma$-factor. We find from Eq.~(\ref{eq:4.12}) that, for the electron under consideration $(R= \gamma - p_z/m_ec = 1)$, the $\gamma$-factor resulting from the motion in the $(x,z)$-plane is $\gamma = 1 + (a_0^2 /2) \sin^2(\omega \tau)$. Here we assumed that $a = a_0 \sin(\xi)$ and took into account that $\xi = \omega \tau$ for $R = 1$. 

Equation (\ref{eq:5.9}) is similar to that of an oscillator with a modulated natural frequency, or a parametric oscillator~\citep{Landau1960}. This modulation is caused by the modulation of the $\gamma$-factor with frequency $2\omega$. It is well known from classical mechanics that the motion of such an oscillator can become unstable for certain values of the natural frequency. We assume that the density $n_0$ in the channel is significantly sub-critical, so that $\omega_p \ll \omega$. This means that for $a_0 \sim 1$ the natural frequency is small compared to the frequency of the modulations. This regime is stable due to the frequency mismatch. The natural frequency increases with wave amplitude due to the $\gamma$-factor multiplier in Eq.~(\ref{eq:5.9}) and it becomes comparable to the frequency of the modulations at $a_0 \sim \omega/\omega_p$. In this regime, the restoring force acting on the electron in the channel changes on a time scale comparable to the period of the oscillations, which provides an opportunity to amplify the amplitude of the oscillations. It is shown by \citet{Arefiev2014} that $a_0 = 1.62  \omega/\omega_p$ is the wave amplitude threshold above which the transverse oscillations become parametrically unstable and their amplitude grows exponentially. It is worth emphasizing that the amplification threshold is determined only by a single combination of parameters $a_0 \omega_p / \omega$.

\begin{figure}
  \subfigure
  {\includegraphics[width=6.5cm]{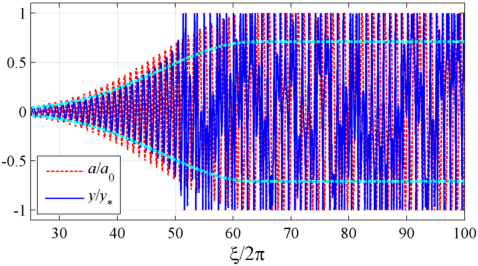}} 
  \quad
  \subfigure
  {\includegraphics[width=6.5cm]{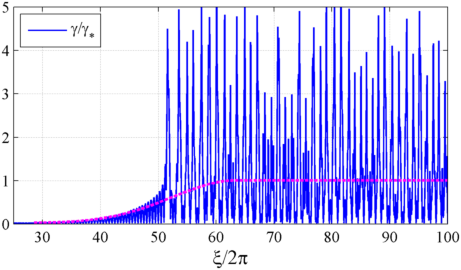}}
\caption{Electron dynamics in the regime where the laser electric field is directed across the channel and the transverse oscillations are unstable ($a_0 = 10$ and $\omega_p/\omega = 0.175$). The dashed (red) curve in the left panel is the amplitude of the laser pulse. The dashed (magenta) curve in the right panel is the maximum $\gamma$-factor in the vacuum regime (without the static field of the channel).} \label{fig:101}
\end{figure}

Figure~\ref{fig:100} shows a solution of the full system of equations [Eqs.~(\ref{Eq1}) - (\ref{Eq4})] for a laser pulse with maximum amplitude $a_0 =10$ and an under-dense ion channel with $\omega_p/\omega = 0.175$. 
For these parameters, the combination $a_0 \omega_p/\omega$ exceeds the threshold for the parametric amplification, $a_0 \omega_p/\omega = 1.75 > 1.62$ . The electron oscillations must become unstable once the laser amplitude exceeds $|a| = 9.3$. The amplitude of oscillations across the channel (see left panel in Fig.~\ref{fig:100}) indeed experiences a rapid increase. The oscillations are quickly amplified to the level comparable to $y_*$ and, at that point, the dephasing $d\xi/d \tau$ becomes significantly reduced. This allows for the electron to stay longer in phase with the wave, leading to a considerable enhancement of the laser-driven acceleration and a resulting enhancement of $\gamma$, as shown in the right panel of Fig.~\ref{fig:100}. For comparison, the dashed (magenta) curve in the right panel of Fig.~\ref{fig:100} shows the maximum $\gamma$-factor for the same pulse in the vacuum regime (no transverse static electric field), as predicted by Eq.~(\ref{eq:4.12})

Next, we consider a regime where the laser electric field is directed across the channel ($\theta = \pi/2$) and, therefore, it can directly drive electron oscillations. Figure~\ref{fig:101} shows the electron dynamics for $a_0 = 10$ and $\omega_p/\omega = 0.1$. In order to distinguish the impact of the channel, we 
also show the vacuum solution (no transverse static electric field) for the same pulse. The dashed curve in the left panel shows the extent of the transverse oscillations, whereas the dashed curve in the right panel shows the maximum $\gamma$-factor in the vacuum regime. Initially, electron oscillations across the channel undergo gradual amplification with the increase of the laser field amplitude. There is a good agreement with the vacuum solution, which indicates that the oscillations are driven by the laser electric field and that the effect of the field of the channel is insignificant at this stage. However, at $\xi/2 \pi > 50$ the oscillations undergo additional amplification and their maximum amplitude becomes close to $y_*$. This leads to a decrease in the dephasing according to Eq.~(\ref{R_main_0}) and, subsequently, to an increase in the electron energy (see right panel of Fig.~\ref{fig:101}). The threshold behavior in Fig.~\ref{fig:101} is similar to that in Fig.~\ref{fig:100}.

In order to determine the extent of the described effect, parameter scans were performed by \citet{Arefiev2014} for several different polarizations by varying $a_0$ and $\omega_p/\omega$ ($1 \leq a_0 \leq 20$ and $0.01 \leq \omega_p/\omega \leq 0.3$). The quantity that was compared is the maximum $\gamma$-factor achieved by the electron for each set of parameters. It was shown that for a given channel density there is indeed a wave amplitude threshold where the maximum $\gamma$-factor has a sharp jump. It is important to point out that the maximum $\gamma$-factor remains enhanced for a wide range of wave amplitudes above the threshold value. The threshold is determined by the combination $a_0 \omega_p/\omega$.

\begin{figure}
  \subfigure
  {\includegraphics[width=6.5cm]{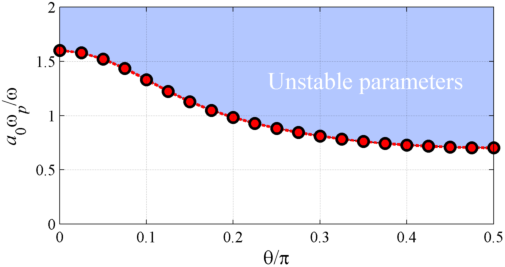}} 
  \quad
  \subfigure
  {\includegraphics[width=6.5cm]{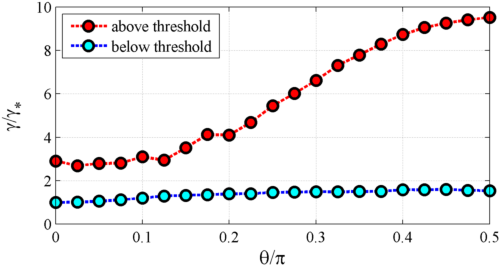}}
\caption{Threshold for the enhancement of $\gamma$ and the corresponding enhancement at the threshold for different laser polarizations specified by the angle $\theta$. The $\gamma$-factor is normalized to $\gamma_* \equiv 1+ a_0^2/2$. The upper curve in the right panel corresponds to the threshold parameters shown with circles in the left panel. The lower curve corresponds to $\omega_p/\omega$ that are smaller than those for the upper curve by just $3\times10^{-4}$.} \label{fig:102}
\end{figure}

Figure~\ref{fig:102} shows how the threshold and the corresponding energy enhancement change with the laser polarization specified by the angle $\theta$. These results were obtained by setting $a_0 = 10$ and repeating calculations similar to those shown in Figs.~\ref{fig:100} and \ref{fig:101} for $0.07 \leq \omega_p/\omega \leq 0.19$ at several different polarization angles marked in Fig.~\ref{fig:102} with circles. For a given $\theta$, the maximum $\gamma$-factor gradually increases with $\omega_p/\omega$ until we reach a threshold value (shown in the left panel of Fig.~\ref{fig:102}) where an increase of $\omega_p/\omega$ by just $3\times10^{-4}$ produces a sharp jump in the maximum $\gamma$-factor (shown in the right panel of Fig.~\ref{fig:102}). By increasing $\theta$ from 0 to $\pi/2$ we increase the amplitude of the driving field across the channel (it increases as $\sin(\theta)$). The results of Fig.~\ref{fig:102} indicate that the threshold enhancement in the electron energy gain occurs regardless of the driving field. However, the presence of the driving field facilitates the onset of the instability in the transverse oscillations across the channel and increases the resulting electron energy gain.

In conclusion, we have shown that a static transverse electric field can have a profound effect on electron acceleration in a laser pulse. For a given ion density in the channel, there is a wave amplitude threshold for amplification of the transverse oscillations across the channel. Above the threshold, the amplitude of the oscillations increases and approaches $\sqrt{2} c / \omega_p$. This leads to a significant reduction in the dephasing rate and subsequent enhancement of the electron energy gain. A driving field across the channel is not necessary, since the  amplification of the transverse oscillations is caused by modulations of the $\gamma$-factor.

%+++++++++++++++++++++++++++++++++++++++++++++++++++++++++

\section{Role of electron injection} \label{Sec6}

In the previous two sections (Secs.~\ref{Sec4} and \ref{Sec5}) we examined how static electric fields impact the dynamics of an electron accelerated by a plane electromagnetic wave. In the analysis, we always assumed that initially the electron is already in the channel, but the laser pulse has not yet reached the electron location longitudinally.  On the other hand, in the case of a steady-state channel discussed in Sec.~\ref{Sec2}, electrons are injected into the channel from the side with the laser beam already present in the channel. This raises a question regarding the role of electron injection. In order to examine transverse electron injection self-consistently, the model formulated in Sec.~\ref{Sec3} would have to be revised to incorporate a laser beam of a finite width as opposed to a plane wave, which goes beyond the scope of this paper. The model of Sec.~\ref{Sec3} however does allow us to examine the role of electron injection in the context of initial conditions. 

Our first step is to determine what part of the electron phase-space is occupied by the solutions presented in Sec.~\ref{Sec5}. We again consider an axially uniform ion channel and, to simplify the discussion, we limit our attention to the case where the laser electric field is directed across the channel. As in Sec.~\ref{Sec5}, we assume that initially the laser pulse has not yet reached the electron location. The electron is initially at rest, located on the axis of the channel. The amplitude of the laser pulse envelope slowly increases to $a_0 = 5$ and then remains constant at $\xi \geq 500$, so that $a(\xi) = a_0 \sin(\xi)$ is a periodic function. In order to visualize the periodicity of the electron motion at $\xi \geq 500$, it is convenient to use a variable that changes periodically from 0 to $2 \pi$, $\psi \equiv \xi - 2\pi\mbox{mod}(\xi,2\pi)$, instead of using $\xi$ that increases monotonically.

\begin{figure}
  \subfigure
  {\includegraphics[width=6cm]{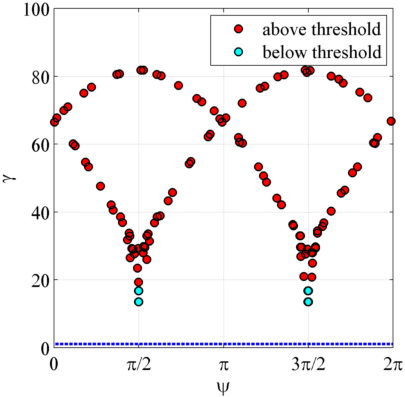}} 
  \quad
  \subfigure
  {\includegraphics[width=7cm]{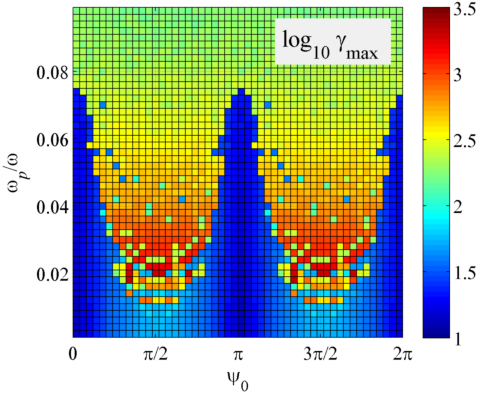}}
\caption{(Left) Electron $\gamma$-factor during crossings of the channel axis and (Right) the maximum $\gamma$-factor [$\log_{10}(\gamma_{\max})$] achieved by injected electrons. The laser amplitude in all the cases is $a_0 = 5$. The data in the left panel is for a gradual laser field ramp-up and $\omega_p/\omega = 0.01$, 0.12 (below the threshold) and $\omega_p/\omega = 0.14$ (above the threshold). The phase $\psi$ is defined as $\psi \equiv \xi - 2\pi\mbox{mod}(\xi,2\pi)$. The data for injected electrons in the right panel corresponds to initial conditions shown on the left with a dashed line. The initial injection phase defines the initial wave amplitude, $a = a_0 \sin(\psi_0)$.} \label{fig:injection_1}
\end{figure}

Figure~\ref{fig:injection_1} shows the electron $\gamma$-factor as a function of $\psi$ at those instants when the electron crosses the axis of the channel for three different channel densities, $\omega_p/\omega = 0.01$, 0.12, and 0.14. According to Fig.~\ref{fig:102}, $\omega_p/\omega = 0.01$ and 0.12 are below the threshold for $a_0 = 5$, whereas $\omega_p/\omega = 0.14$ is above the threshold. For the initial conditions that we are considering, the electron is initially located on axis with $\gamma = 1$ and $\psi = 0$. The oscillating electric field of the laser pulse drives electron oscillations across the channel. As the envelope of the laser pulse increases from 0 to $a_0$, the electron $\gamma$-factor at the instances when the electron crosses the axis increases as well. We do not show this transition in Fig.~\ref{fig:injection_1}, focusing only on the regime after the envelope has reached its constant amplitude ($\xi \geq 500$). Figure~\ref{fig:injection_1} shows that the electron motion is periodic below the threshold. The electron crosses the axis twice every wave period at $\psi = \pi/2$ and $3 \pi/2$ with the same value of $\gamma$ (lower cyan circles for $\omega_p/\omega = 0.01$ and upper cyan circles for $\omega_p/\omega = 0.12$). The situation changes dramatically above the threshold (red circles), as the timing of the electron crossing is no longer constant and the corresponding $\gamma$-factor increases substantially. The results are shown for $500 \leq \xi \leq 700$. Electron motion becomes chaotic at $\xi \gg 700$, so that the pattern visible in Fig.~\ref{fig:injection_1} starts to disappear. 

We can therefore conclude that the initial conditions used in Sec.~\ref{Sec5} and corresponding to the laser field being turned on gradually cover only a very limited area of the electron parameter space if $\omega_p/\omega$ is below the threshold for amplification of transverse oscillations. Specifically, the electron crosses the axis of the channel only at $\psi = \pi/2$ and $3 \pi/2$ and with the same $\gamma$ for a given $\omega_p/\omega$.

Our next step is to examine the impact of electron injection into a laser pulse on electron dynamics and maximum energy gain. We consider an electron that is injected into a laser beam with $a = a_0 \sin(\xi + \psi_0)$, where $\psi_0$ is the phase of the laser pulse at the moment of injection that we define as $\xi = 0$. We are particularly interested in those regimes where $\omega_p/\omega$ is below the threshold determined in Sec.~\ref{Sec5}, so that the electron motion is stable and there is no significant energy enhancement if the laser pulse is turned on gradually. We deliberately consider electrons that are injected with initial parameters that cannot be achieved by gradually ramping up the laser pulse. For simplicity, we assume that electrons are injected on the axis without any momentum. This way there is only one free parameter, which is the injection phase $\psi_0$. These initial conditions are shown with a dashed line in the left panel of Fig.~\ref{fig:injection_1}. Clearly, an electron in the previously considered regime is unable to achieve any of these states while moving in a laser beam of full amplitude.

The right panel of Fig.~\ref{fig:injection_1} shows how the maximum $\gamma$-factor changes for the injected electrons depending on the injection phase $\psi_0$ and the ion density that determines $\omega_p/\omega$. The maximum $\gamma$-factor is color-coded and shown on a log-scale, $\log_{10}(\gamma_{\max})$. At $\omega_p/\omega = 0.01$, the electron motion is regular for all injection phases. The maximum $\gamma$-factor peaks for $\psi_0 = \pi/2$ and $3\pi/2$ at $\gamma_{\max} \approx 54$, and has the lowest value of $\gamma_{\max} \approx 14$ at $\psi_0 = 0$ and $\pi$. This is an effect of the initial injection phase that takes place in a purely vacuum case as well. Indeed, repeating the steps of Sec.~\ref{Sec4} for 
$a = a_0 \sin(\xi+\psi_0)$ and an electron without any initial momentum, we find that $p_y/m_e c = a_0 [\sin(\xi + \psi_0) - \sin(\psi_0)]$ and, as a result, $\gamma = 1+ a_0^2 [\sin(\xi + \psi_0) - \sin(\psi_0)]^2/2$. The maximum $\gamma$-factor is then roughly four times higher for $\psi_0 = \pi/2$ and $3\pi/2$ than for $\psi_0 = 0$ and $\pi$ because the maximum transverse momentum is doubled. 

\begin{figure}
  \subfigure
  {\includegraphics[width=7cm]{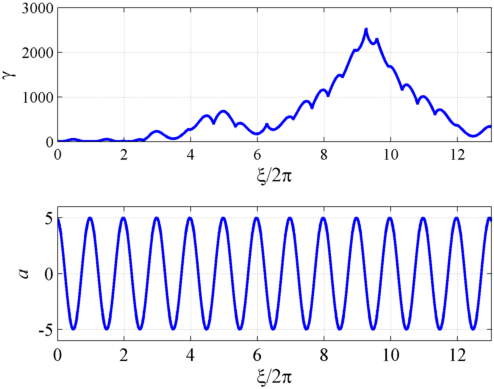}} 
  \quad
  \subfigure
  {\includegraphics[width=6cm]{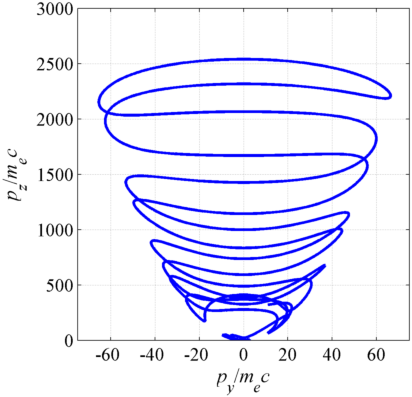}}
\caption{Dynamics of an electron injected without initial momentum into 
a wave with $a(\xi) = a_0 \sin(\xi + \psi_0)$, where $a_0 = 5$ and $\psi = 0.56 \pi$. The electron is injected onto the axis of the channel where $\omega_p/\omega = 0.016$.} \label{fig:injection_2}
\end{figure}

In can be seen in the right panel of Fig.~\ref{fig:injection_1} that the injection phases $\psi_0 = \pi/2$ and $3\pi/2$ are the most unstable ones. At $\omega_p/\omega \approx 0.015$ and above, the electron motion across the channel becomes irregular for these phases as it undergoes amplification. The maximum $\gamma$-factor that the electron achieves becomes significantly increased as well. On the other hand, $\psi_0 = 0$ and $\pi$ are the least unstable injection phases. The threshold for the onset of the irregular motion and a resulting electron energy increase has apparently a strong dependence on the injection phase. It is worth pointing out that even at $\psi_0 = 0$ and $\pi$ the threshold is still well below what we would expect from Fig.~\ref{fig:102}.

The highest $\gamma$-factor is achieved by electrons injected close to zeros of the laser field ($\psi_0 = \pi/2$ and $3\pi/2$). An example of electron dynamics in this regime is shown in Fig.~\ref{fig:injection_2}, where $\psi_0 \approx 0.56 \pi$ and $\omega_p/\omega = 0.016$. The left two panels in Fig.~\ref{fig:injection_2} show that the energy gain is a continuous process that occurs during multiple oscillations of the laser field. The right panel shows the electron trajectory in the momentum space. The electron longitudinal momentum continues to increase as the electron makes multiple oscillations across the channel (evident from the change of sign of $p_y$), which is a clear indicator  that the longitudinal force remains positive for extended periods of time. The remarkable prolonged acceleration is achieved due to the fact that $p_y$ and $B_{wave}$ remain in phase despite their oscillations. In other words, the two can remain in phase while the phase $\xi$ of the wave at the electron location increases by much more than $2 \pi$. The eventual dephasing between $p_y$ and $B_{wave}$ limits the maximum electron energy gain. 

In conclusion, we have shown that a slow laser pulse ramp-up significantly limits the parameter space accessible to the electron if the ion density is below the threshold established in Sec.~\ref{Sec5}. Electron injection into the laser pulse can significantly lower the threshold for the electron energy enhancement. The injection also enables novel regimes where extended longitudinal acceleration is possible due to slow dephasing between the transverse electron momentum and the magnetic field of the wave.

%+++++++++++++++++++++++++++++++++++++++++++++++++++++++++

\section{Role of a static longitudinal magnetic field} \label{Sec7}

Recent developments \citep{Courtois2005,Daido1986,Fujioka2012,Fujioka2013,Albertazzi2014} in the generation of strong magnetic fields on large scales using high-power laser techniques raise the question as to what use might be made of magnetic fields with flux densities that perhaps even reach 1kT. Theoretical work also continues to indicate that self-generated longitudinal magnetic fields can occur in laser-plasma interactions \citep{Liu2013}. Both of these are good reasons to re-examine the fundamental role that longitudinal magnetic fields play in direct laser acceleration of electrons in ultra-intense laser-plasma interactions, as one may have either a self-generated axial magnetic field or one may wish to contemplate experiments in which an externally generated axial magnetic field is imposed.

In this section, we employ the single electron model of Sec.~\ref{Sec3} to examine the role of a static magnetic field. We consider a setup where a uniform static magnetic field $B_0$ is directed longitudinally along the laser pulse propagation. In order to simplify the problem, we neglect transverse static electric fields, but retain a weak static longitudinal electric field $E_0$ that is present in a limited region along the $z$-axis. Without any loss of generality, we assume that the laser electric field is polarized along the $x$-axis [which corresponds to $\theta = 0$ in Eq.~(\ref{eq:a})]. It then follows from Eq.~(\ref{main_eq:2}) that the evolution of the electron momentum is described by the following coupled equations:  
\begin{eqnarray}
&& \frac{1}{\omega} \frac{d}{d \tau} \left( \frac{p_x}{m_e c} - a \right) = - \frac{\omega_{ce}}{\omega} \frac{p_y}{m_e c}, \label{7_main:1} \\
&& \frac{1}{\omega} \frac{d}{d \tau} \left( \frac{p_y}{m_e c} \right) = \frac{\omega_{ce}}{\omega} \frac{p_x}{m_e c}, \label{7_main:2} \\
&& \frac{1}{\omega} \frac{d}{d \tau} \left( \frac{p_z}{m_e c} \right)  = - \gamma \frac{|e| E_0}{m_e c \omega} + \frac{p_x}{m_e c} \frac{da}{d \xi}, \label{7_main:3}
\end{eqnarray}
where $\omega_{ce} \equiv |e|B_0/m_ec$ is the classical electron cyclotron frequency and $\tau$ is the proper time defined by Eq.~(\ref{eq:4.5}). Here we took into account the expressions for the wave electric and magnetic fields given by Eqs.~(\ref{wave:1}) and (\ref{wave:2}). The electron coordinates evolve according to Eq.~(\ref{main_eq:1}), whereas the phase $\xi$ evolves according to Eq.~(\ref{eq:deph_2}),
\begin{equation}
\frac{1}{\omega} \frac{d \xi}{d \tau} = \gamma - p_z/m_ec.
\end{equation}

\begin{figure}
  \subfigure
  {\includegraphics[width=6.5cm]{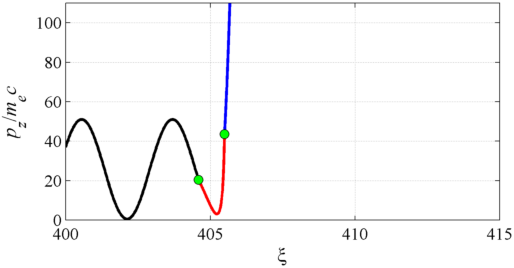}} 
  \quad
  \subfigure
  {\includegraphics[width=6.5cm]{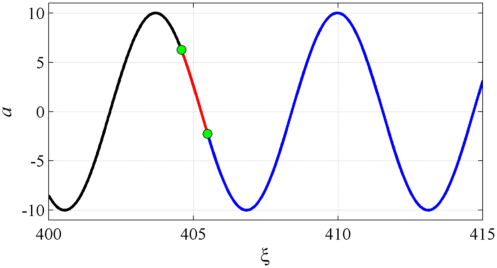}}

  \subfigure
  {\includegraphics[width=6.5cm]{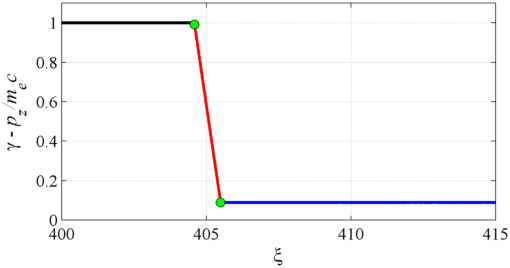}} 
  \quad
  \subfigure
  {\includegraphics[width=6.5cm]{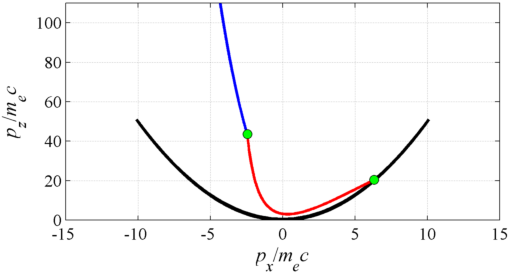}}
\caption{Dynamics of an initially immobile electron irradiated by a plane wave with $a_0=10$ in a static longitudinal magnetic field and passing through a region with a longitudinal electric field $|e| E_0 /m_e \omega c = - 0.1 a_0$ located at $2413 \leq z/\lambda \leq 2445$. The dynamics before, during, and after the interaction with $E_0$ is shown in black, red, and blue. The ratio of the cyclotron to wave frequency is $\omega_{ce}/\omega = 0.1$.} \label{fig:cycl_1}
\end{figure}

It follows directly from Eqs.~(\ref{7_main:1}) - (\ref{7_main:3}) that
\begin{equation} \label{7_main:4}
\frac{d}{d \tau} \left( \gamma - \frac{p_z}{m_e c}\right) = \frac{|e| E_0}{\omega m_ e c} \frac{d \xi}{d \tau}.
\end{equation}
This relation is identical to the one that was derived in Sec.~\ref{Sec4}. We then conclude that in the absence of the longitudinal field the dephasing rate $R = \gamma - p_z/m_ec$ remains constant, so that
\begin{equation}
\frac{1}{\omega} \frac{d \xi}{d \tau} = R,
\end{equation}
where $R$ is determined by the initial conditions. Taking this into account, we can combine Eqs.~(\ref{7_main:1}) and (\ref{7_main:2}) to obtain a fully self-contained equation for $p_y$,
\begin{equation} \label{eq:7.7}
\frac{d^2 p_y}{d \xi^2} + \left( \frac{\omega_{ce}}{R \omega} \right)^2 p_y = \frac{\omega_{ce}}{\omega} \frac{m_ec}{R} \frac{da}{d \xi}.
\end{equation}
Equation (\ref{eq:7.7}) is an equation for a driven harmonic oscillator with natural frequency $\omega_{ce}/R \omega$. The frequency of the driving force in this equation is unity, since $a \propto \sin(\xi)$. It is instructive to compare Eq. (\ref{eq:7.7}) to Eq.~(\ref{eq:5.9}) for transverse electron oscillations across the ion channel. There is no $\gamma$-factor multiplier in Eq.~(\ref{eq:7.7}) that allows for the natural frequency of the oscillations in Eq.~(\ref{eq:5.9}) to increase as a result of electron acceleration. As a result, if there is an initial mismatch between the natural frequency and the frequency of the driver in Eq.~(\ref{eq:7.7}), then it will persist throughout electron acceleration. 

A resonance condition in the context of Eq.~(\ref{eq:7.7}) is when the natural frequency of the oscillations is equal to unity. Let us first consider an electron that is initially at rest, so that $R = 1$. The resonance condition for this electron is $\omega_{ce}/\omega =1$. For a laser pulse with wavelength $\lambda = 1$ $\mu$m, the cyclotron frequency will be ten times smaller than the frequency of the laser even in the case of a 1kT magnetic field. This implies that the potential for significant electron cyclotron resonance absorption is rather limited.

Our assessment of the role of electron cyclotron acceleration might be changed by considering either pre-acceleration of electrons (i.e. initially having longitudinal momentum) or the acceleration in a longitudinal electric field.  
Longitudinal pre-acceleration lowers the value of the constant of motion $R = \gamma-p_z/m_ec$, effectively increasing the natural frequency of oscillations [see Eq.~(\ref{eq:7.7})]. In the limit of $g_z \gg m_ec$ and $g_x = g_y = 0$, where $g$ is the initial electron momentum, one has $R \approx m_e c / 2 g_z$. Clearly, the natural frequency can be substantially increased by boosting the initial longitudinal momentum to relativistic values, while it can still remain much less than $a_0^2/2$. This means that pre-acceleration plays a somewhat subtle role in the cyclotron configuration, increasing the effective cyclotron frequency by reducing the longitudinal dephasing between the electron and the laser pulse.

\begin{figure}
  \subfigure
  {\includegraphics[width=6.5cm]{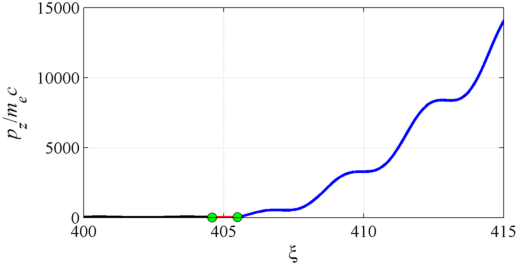}} 
  \quad
  \subfigure
  {\includegraphics[width=6.0cm]{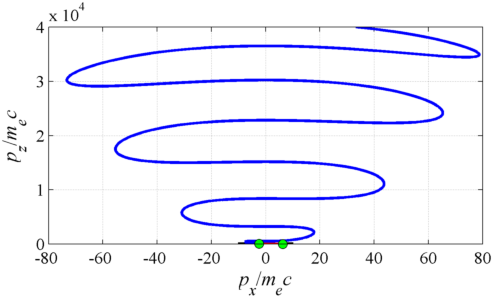}}
\caption{Evolution of electron momentum for the same set of parameters as in Fig.~\ref{fig:cycl_1} plotted using a different vertical scale to show the scale of the electron energy gain following the interaction with the longitudinal field.} \label{fig:cycl_2}
\end{figure}

The same effect can be achieved by employing a negative longitudinal electric field. As indicated by Eq.~(\ref{7_main:4}), this field would lead to a reduction of the dephasing rate. The mechanism is the same as the one discussed in detail in Sec.~\ref{Sec4}. Therefore, a combination of longitudinal magnetic and electric fields can be extremely beneficial for increasing the electron energy gain from the laser pulse. An example is shown in Figs.~\ref{fig:cycl_1} and \ref{fig:cycl_2}, where an initially immobile electron is irradiated by a laser pulse with $a_0 = 10$ in a strong longitudinal magnetic field with $\omega_{ce} = 0.1 \omega$. As the electron accelerates and starts to move in the direction of the laser pulse propagation, it encounters a region with a relatively weak longitudinal electric field, $|e| E_0 /m_e \omega c = - 0.1 a_0$. The interaction with the field $E_0$ takes place near a zero of the laser electric field, which allows for the maximum reduction of the dephasing rate (see Fig.~\ref{fig:cycl_1}). The dephasing rate drops from $R=1$ to $R \approx 0.089$, which leads to a frequency match between the transverse electron oscillations and the oscillations of the laser field, with $\omega_{ce}/R \omega \approx 1.12$. The resonant interaction with the laser field that follows the interaction with the longitudinal electric field is evident in Fig.~\ref{fig:cycl_2}. The amplitude of the oscillating transverse electron momentum $p_x$ increases, which leads to an increased longitudinal force and, consequently, to an enhancement of the longitudinal momentum. A positive longitudinal electric field, on the other hand, can have the reverse effect, and can strongly pull electrons away from resonance. So when analyzing interactions, the detailed structure of the longitudinal electric field becomes critical.

In conclusion, we have re-examined the effect of a static longitudinal magnetic field on electron motion and we have shown that a resonance condition between the electron cyclotron motion and the laser field remains unaffected by electron acceleration in the laser pulse. This aspect qualitatively distinguishes the impact of a longitudinal magnetic field from that of a static electric field on transverse electron motion.  If an electron is initially far from resonance then it remains so. However, pre-acceleration of the electron or acceleration by a longitudinal electric field can reduce the dephasing rate of the electron and thus increase the effective electron cyclotron frequency, a point that was not made in prior examinations of this configuration, e.g. \citep{Liu2004,Tsakiris2000}. This can then lead to a resonant interaction with the laser field, enabling a considerable electron energy gain. Deacceleration by a longitudinal electric field can lower the effective electron cyclotron frequency, moving the electron away from resonance. Therefore, the consideration of the longitudinal electric field is an essential part of any interaction with a significant axial magnetic field.

%+++++++++++++++++++++++++++++++++++++++++++++++++++++++++

\section{Summary and discussion} \label{Sec8}

We have examined the impact of several factors on electron acceleration by a laser pulse and the resulting electron energy gain. Specifically, we have considered the role played by: 1) static longitudinal electric field; 2) static transverse electric field; 3) electron injection into the laser pulse; and 4) static longitudinal magnetic field. It is shown that all of these factors can affect the electron dynamics, leading, under certain conditions, to a considerable electron energy gain. 

We have shown that even a relatively weak localized static longitudinal electric field strongly affects the electron dynamics by changing the dephasing rate. A field directed against the laser pulse propagation can significantly decrease the dephasing rate without having to transfer a considerable amount of energy to the electron during the interaction. A considerable energy gain takes place after the interaction, with the extra energy being transferred to the electron from the laser pulse.

A static transverse electric field can also have a profound effect of electron acceleration in a laser pulse. For a given ion density in the channel that generates the field, there is a wave amplitude threshold for amplification of the transverse oscillations across the channel. Above the threshold, the amplitude of the oscillations increases and approaches $\sqrt{2} c / \omega_p$. This leads to a significant reduction in the dephasing rate and subsequent enhancement of the electron energy gain. A driving field across the channel is not necessary (yet beneficial) to achieve an energy gain, since the amplification of the transverse oscillations is caused by modulations of the $\gamma$-factor.

We have also found that a slow laser pulse ramp-up significantly constrains the parameter space accessible to an electron irradiated by such a pulse in a plasma channel. Electron injection into the laser pulse can significantly lower the threshold for the electron energy enhancement and it can also enable novel regimes where extended longitudinal acceleration is possible due to slow dephasing between the transverse electron momentum and the magnetic field of the wave.

We have also re-examined the effect of a static longitudinal magnetic field on electron motion and we have shown that a resonance condition between the electron cyclotron motion and the laser field remains unaffected by electron acceleration in the laser pulse. This aspect qualitatively distinguishes the impact of a longitudinal magnetic field from that of a static electric field on transverse electron motion. We have shown that pre-acceleration of the electron or acceleration by a longitudinal electric field can be extremely beneficial for achieving the resonance. This is because a reduction of the dephasing rate increases the effective electron cyclotron frequency, which can lead to a resonant frequency match and thus enable a considerable electron energy gain.

The mechanisms described in this paper open the possibility for enhancement and control of the electron energy gain in laser-plasma interactions by designing an appropriate field and plasma configuration. Self-consistent 3D modeling that incorporates ion dynamics, finite width of the laser pulse, and the effect of the plasma on the laser pulse propagation is required to make quantitative predictions based on the proposed mechanisms. Nevertheless, the robust effects outlined in this work already allow one to make qualitative predictions. For example, the results of
Sec.~\ref{Sec5} predict a sharp density threshold for generation of energetic electrons. A very similar feature, including a gradual decrease in characteristic electron energies above the threshold [see lower panel of Fig. 8 in \citep{Arefiev2014}], was observed by \citet{Mangles2005} in experiments with gas jets. Finally, more analysis is required in order to understand the impact that surface waves that develop in long plasma channels \citep{Willingale2013} can have on electron injection and subsequent acceleration by the laser beam. 

%+++++++++++++++++++++++++++++++++++++++++++++++++++++++++

\section{Acknowledgments}

AVA would like to thank Dr. Louise Willingale and Dr. Toma Toncian for constructive comments and David Stark for editing the manuscript. Simulations for this paper were performed using the EPOCH code (developed under UK EPSRC grants EP/G054940/1, EP/G055165/1 and EP/G056803/1) using HPC resources provided by the Texas Advanced Computing Center at The University of Texas. AVA was supported by AFOSR Contract No. FA9550-14-1-0045, National Nuclear Security Administration Contract No. DE-FC52-08NA28512 and U.S. Department of Energy Contract No. DE-FG02-04ER54742. VNK was supported by AFOSR Contract No. FA9550-14-1-0045 and U.S. Department of Energy Contracts No. DE-SC0007889 and DE-SC0010622.

%+++++++++++++++++++++++++++++++++++++++++++++++++++++++++

%\bibliographystyle{jpp}

\bibliography{arxiv_2015}

%Merlin.mbs v4.21 2009-07-09.
\begin{thebibliography}{10}%
\makeatletter
\providecommand \@ifxundefined [1]{%
 \ifx #1\undefined \expandafter \@firstoftwo
 \else \expandafter \@secondoftwo
\fi
}%
\providecommand \@ifnum [1]{%
 \ifnum #1\expandafter \@firstoftwo
 \else \expandafter \@secondoftwo
\fi
}%
\providecommand \enquote [1]{``#1''}%
\providecommand \bibnamefont  [1]{#1}%
\providecommand \bibfnamefont [1]{#1}%
\providecommand \citenamefont [1]{#1}%
\providecommand\href[0]{\@sanitize\@href}%
\providecommand\@href[1]{\endgroup\@@startlink{#1}\endgroup\@@href}%
\providecommand\@@href[1]{#1\@@endlink}%
\providecommand \@sanitize [0]{\begingroup\catcode`\&12\catcode`\#12\relax}%
\@ifxundefined \pdfoutput {\@firstoftwo}{%
 \@ifnum{\z@=\pdfoutput}{\@firstoftwo}{\@secondoftwo}%
}{%
 \providecommand\@@startlink[1]{\leavevmode\special{html:<a href="#1">}}%
 \providecommand\@@endlink[0]{\special{html:</a>}}%
}{%
 \providecommand\@@startlink[1]{%
  \leavevmode
  \pdfstartlink
   attr{/Border[0 0 1 ]/H/I/C[0 1 1]}%
   user{/Subtype/Link/A<</Type/Action/S/URI/URI(#1)>>}%
  \relax
 }%
 \providecommand\@@endlink[0]{\pdfendlink}%
}%
\providecommand \url  [0]{\begingroup\@sanitize \@url }%
\providecommand \@url [1]{\endgroup\@href {#1}{\urlprefix}}%
\providecommand \urlprefix [0]{URL }%
\providecommand \Eprint[0]{\href }%
\@ifxundefined \urlstyle {%
  \providecommand \doi [1]{doi:\discretionary{}{}{}#1}%
}{%
  \providecommand \doi [0]{doi:\discretionary{}{}{}\begingroup
  \urlstyle{rm}\Url }%
}%
\providecommand \doibase [0]{http://dx.doi.org/}%
\providecommand \Doi[1]{\href{\doibase#1}}%
\providecommand \bibAnnote [3]{%
  \BibitemShut{#1}%
  \begin{quotation}\noindent
    \textsc{Key:}\ #2\\\textsc{Annotation:}\ #3%
  \end{quotation}%
}%
\providecommand \bibAnnoteFile [2]{%
  \IfFileExists{#2}{\bibAnnote {#1} {#2} {\input{#2}}}{}%
}%
\providecommand \typeout [0]{\immediate \write \m@ne }%
\providecommand \selectlanguage [0]{\@gobble}%
\providecommand \bibinfo [0]{\@secondoftwo}%
\providecommand \bibfield [0]{\@secondoftwo}%
\providecommand \translation [1]{[#1]}%
\providecommand \BibitemOpen[0]{}%
\providecommand \bibitemStop [0]{}%
\providecommand \bibitemNoStop [0]{.\EOS\space}%
\providecommand \EOS [0]{\spacefactor3000\relax}%
\providecommand \BibitemShut [1]{\csname bibitem#1\endcsname}%
%</preamble>
\bibitem{Flippo2010}%
  \BibitemOpen
  \bibfield{author}{%
  \bibinfo {author} {\bibfnamefont{K.}~\bibnamefont{Flippo}}, \bibinfo {author}
  {\bibfnamefont{T.}~\bibnamefont{Bartal}}, \bibinfo {author}
  {\bibfnamefont{F.}~\bibnamefont{Beg}}, \bibinfo {author}
  {\bibfnamefont{S.}~\bibnamefont{Chawla}}, \bibinfo {author}
  {\bibfnamefont{J.}~\bibnamefont{Cobble}}, \bibinfo {author}
  {\bibfnamefont{S.}~\bibnamefont{Gaillard}}, \bibinfo {author}
  {\bibfnamefont{D.}~\bibnamefont{Hey}}, \bibinfo {author}
  {\bibfnamefont{A.}~\bibnamefont{MacKinnon}}, \bibinfo {author}
  {\bibfnamefont{A.}~\bibnamefont{MacPhee}}, \bibinfo {author}
  {\bibfnamefont{P.}~\bibnamefont{Nilson}}, \bibinfo {author}
  {\bibfnamefont{D.}~\bibnamefont{Offermann}}, \bibinfo {author}
  {\bibfnamefont{S.}~\bibnamefont{Le~Pape}},\ and\ \bibinfo {author}
  {\bibfnamefont{M.~J.}\ \bibnamefont{Schmitt}},\ }%
  \bibfield{journal}{%
  \bibinfo {journal} {J. Phys.: Conf. Ser.}\ }%
  \textbf{\bibinfo {volume} {244}},\ \bibinfo {pages} {022033} (\bibinfo {year}
  {2010})%
  \bibAnnoteFile{NoStop}{Flippo2010}%
\bibitem{Kneip2008}%
  \BibitemOpen
  \bibfield{author}{%
  \bibinfo {author} {\bibfnamefont{S.}~\bibnamefont{Kneip}}, \bibinfo {author}
  {\bibfnamefont{S.~R.}\ \bibnamefont{Nagel}}, \bibinfo {author}
  {\bibfnamefont{C.}~\bibnamefont{Bellei}}, \bibinfo {author}
  {\bibfnamefont{N.}~\bibnamefont{Bourgeois}}, \bibinfo {author}
  {\bibfnamefont{A.~E.}\ \bibnamefont{Dangor}}, \bibinfo {author}
  {\bibfnamefont{A.}~\bibnamefont{Gopal}}, \bibinfo {author}
  {\bibfnamefont{R.}~\bibnamefont{Heathcote}}, \bibinfo {author}
  {\bibfnamefont{S.~P.~D.}\ \bibnamefont{Mangles}}, \bibinfo {author}
  {\bibfnamefont{J.~R.}\ \bibnamefont{Marqu\`es}}, \bibinfo {author}
  {\bibfnamefont{A.}~\bibnamefont{Maksimchuk}}, \bibinfo {author}
  {\bibfnamefont{P.~M.}\ \bibnamefont{Nilson}}, \bibinfo {author}
  {\bibfnamefont{K.~T.}\ \bibnamefont{Phuoc}}, \bibinfo {author}
  {\bibfnamefont{S.}~\bibnamefont{Reed}}, \bibinfo {author}
  {\bibfnamefont{M.}~\bibnamefont{Tzoufras}}, \bibinfo {author}
  {\bibfnamefont{F.~S.}\ \bibnamefont{Tsung}}, \bibinfo {author}
  {\bibfnamefont{L.}~\bibnamefont{Willingale}}, \bibinfo {author}
  {\bibfnamefont{W.~B.}\ \bibnamefont{Mori}}, \bibinfo {author}
  {\bibfnamefont{A.}~\bibnamefont{Rousse}}, \bibinfo {author}
  {\bibfnamefont{K.}~\bibnamefont{Krushelnick}},\ and\ \bibinfo {author}
  {\bibfnamefont{Z.}~\bibnamefont{Najmudin}},\ }%
  \bibfield{journal}{%
  \bibinfo {journal} {Phys. Rev. Lett.}\ }%
  \textbf{\bibinfo {volume} {100}},\ \bibinfo {pages} {105006} (\bibinfo
  {month} {Mar}\ \bibinfo {year} {2008})%
  \bibAnnoteFile{NoStop}{Kneip2008}%
\bibitem{Chen2010}%
  \BibitemOpen
  \bibfield{author}{%
  \bibinfo {author} {\bibfnamefont{H.}~\bibnamefont{Chen}}, \bibinfo {author}
  {\bibfnamefont{S.~C.}\ \bibnamefont{Wilks}}, \bibinfo {author}
  {\bibfnamefont{D.~D.}\ \bibnamefont{Meyerhofer}}, \bibinfo {author}
  {\bibfnamefont{J.}~\bibnamefont{Bonlie}}, \bibinfo {author}
  {\bibfnamefont{C.~D.}\ \bibnamefont{Chen}}, \bibinfo {author}
  {\bibfnamefont{S.~N.}\ \bibnamefont{Chen}}, \bibinfo {author}
  {\bibfnamefont{C.}~\bibnamefont{Courtois}}, \bibinfo {author}
  {\bibfnamefont{L.}~\bibnamefont{Elberson}}, \bibinfo {author}
  {\bibfnamefont{G.}~\bibnamefont{Gregori}}, \bibinfo {author}
  {\bibfnamefont{W.}~\bibnamefont{Kruer}}, \bibinfo {author}
  {\bibfnamefont{O.}~\bibnamefont{Landoas}}, \bibinfo {author}
  {\bibfnamefont{J.}~\bibnamefont{Mithen}}, \bibinfo {author}
  {\bibfnamefont{J.}~\bibnamefont{Myatt}}, \bibinfo {author}
  {\bibfnamefont{C.~D.}\ \bibnamefont{Murphy}}, \bibinfo {author}
  {\bibfnamefont{P.}~\bibnamefont{Nilson}}, \bibinfo {author}
  {\bibfnamefont{D.}~\bibnamefont{Price}}, \bibinfo {author}
  {\bibfnamefont{M.}~\bibnamefont{Schneider}}, \bibinfo {author}
  {\bibfnamefont{R.}~\bibnamefont{Shepherd}}, \bibinfo {author}
  {\bibfnamefont{C.}~\bibnamefont{Stoeckl}}, \bibinfo {author}
  {\bibfnamefont{M.}~\bibnamefont{Tabak}}, \bibinfo {author}
  {\bibfnamefont{R.}~\bibnamefont{Tommasini}},\ and\ \bibinfo {author}
  {\bibfnamefont{P.}~\bibnamefont{Beiersdorfer}},\ }%
  \bibfield{journal}{%
  \bibinfo {journal} {Phys. Rev. Lett.}\ }%
  \textbf{\bibinfo {volume} {105}},\ \bibinfo {pages} {015003} (\bibinfo
  {month} {Jul}\ \bibinfo {year} {2010})%
  \bibAnnoteFile{NoStop}{Chen2010}%
\bibitem{Pomerantz2014}%
  \BibitemOpen
  \bibfield{author}{%
  \bibinfo {author} {\bibfnamefont{I.}~\bibnamefont{Pomerantz}}, \bibinfo
  {author} {\bibfnamefont{E.}~\bibnamefont{McCary}}, \bibinfo {author}
  {\bibfnamefont{A.~R.}\ \bibnamefont{Meadows}}, \bibinfo {author}
  {\bibfnamefont{A.}~\bibnamefont{Arefiev}}, \bibinfo {author}
  {\bibfnamefont{A.~C.}\ \bibnamefont{Bernstein}}, \bibinfo {author}
  {\bibfnamefont{C.}~\bibnamefont{Chester}}, \bibinfo {author}
  {\bibfnamefont{J.}~\bibnamefont{Cortez}}, \bibinfo {author}
  {\bibfnamefont{M.~E.}\ \bibnamefont{Donovan}}, \bibinfo {author}
  {\bibfnamefont{G.}~\bibnamefont{Dyer}}, \bibinfo {author}
  {\bibfnamefont{E.~W.}\ \bibnamefont{Gaul}}, \bibinfo {author}
  {\bibfnamefont{D.}~\bibnamefont{Hamilton}}, \bibinfo {author}
  {\bibfnamefont{D.}~\bibnamefont{Kuk}}, \bibinfo {author}
  {\bibfnamefont{A.~C.}\ \bibnamefont{Lestrade}}, \bibinfo {author}
  {\bibfnamefont{C.}~\bibnamefont{Wang}}, \bibinfo {author}
  {\bibfnamefont{T.}~\bibnamefont{Ditmire}},\ and\ \bibinfo {author}
  {\bibfnamefont{B.~M.}\ \bibnamefont{Hegelich}},\ }%
  \bibfield{journal}{%
  \bibinfo {journal} {Phys. Rev. Lett.}\ }%
  \textbf{\bibinfo {volume} {113}},\ \bibinfo {pages} {184801} (\bibinfo
  {month} {Oct}\ \bibinfo {year} {2014})%
  \bibAnnoteFile{NoStop}{Pomerantz2014}%
\bibitem{Willingale2013}%
  \BibitemOpen
  \bibfield{author}{%
  \bibinfo {author} {\bibfnamefont{L.}~\bibnamefont{Willingale}}, \bibinfo
  {author} {\bibfnamefont{A.~G.~R.}\ \bibnamefont{Thomas}}, \bibinfo {author}
  {\bibfnamefont{P.~M.}\ \bibnamefont{Nilson}}, \bibinfo {author}
  {\bibfnamefont{H.}~\bibnamefont{Chen}}, \bibinfo {author}
  {\bibfnamefont{J.}~\bibnamefont{Cobble}}, \bibinfo {author}
  {\bibfnamefont{R.~S.}\ \bibnamefont{Craxton}}, \bibinfo {author}
  {\bibfnamefont{A.}~\bibnamefont{Maksimchuk}}, \bibinfo {author}
  {\bibfnamefont{P.~A.}\ \bibnamefont{Norreys}}, \bibinfo {author}
  {\bibfnamefont{T.~C.}\ \bibnamefont{Sangster}}, \bibinfo {author}
  {\bibfnamefont{R.~H.~H.}\ \bibnamefont{Scott}}, \bibinfo {author}
  {\bibfnamefont{C.}~\bibnamefont{Stoeckl}}, \bibinfo {author}
  {\bibfnamefont{C.}~\bibnamefont{Zulick}},\ and\ \bibinfo {author}
  {\bibfnamefont{.~K.}\ \bibnamefont{Krushelnick}},\ }%
  \bibfield{journal}{%
  \bibinfo {journal} {New Journal of Physics}\ }%
  \textbf{\bibinfo {volume} {15}},\ \bibinfo {pages} {025023} (\bibinfo {year}
  {2013})%
  \bibAnnoteFile{NoStop}{Willingale2013}%
\bibitem{Mangles2005}%
  \BibitemOpen
  \bibfield{author}{%
  \bibinfo {author} {\bibfnamefont{S.~P.~D.}\ \bibnamefont{Mangles}}, \bibinfo
  {author} {\bibfnamefont{B.~R.}\ \bibnamefont{Walton}}, \bibinfo {author}
  {\bibfnamefont{M.}~\bibnamefont{Tzoufras}}, \bibinfo {author}
  {\bibfnamefont{Z.}~\bibnamefont{Najmudin}}, \bibinfo {author}
  {\bibfnamefont{R.~J.}\ \bibnamefont{Clarke}}, \bibinfo {author}
  {\bibfnamefont{A.~E.}\ \bibnamefont{Dangor}}, \bibinfo {author}
  {\bibfnamefont{R.~G.}\ \bibnamefont{Evans}}, \bibinfo {author}
  {\bibfnamefont{S.}~\bibnamefont{Fritzler}}, \bibinfo {author}
  {\bibfnamefont{A.}~\bibnamefont{Gopal}}, \bibinfo {author}
  {\bibfnamefont{C.}~\bibnamefont{Hernandez-Gomez}}, \bibinfo {author}
  {\bibfnamefont{W.~B.}\ \bibnamefont{Mori}}, \bibinfo {author}
  {\bibfnamefont{W.}~\bibnamefont{Rozmus}}, \bibinfo {author}
  {\bibfnamefont{M.}~\bibnamefont{Tatarakis}}, \bibinfo {author}
  {\bibfnamefont{A.~G.~R.}\ \bibnamefont{Thomas}}, \bibinfo {author}
  {\bibfnamefont{F.~S.}\ \bibnamefont{Tsung}}, \bibinfo {author}
  {\bibfnamefont{M.~S.}\ \bibnamefont{Wei}},\ and\ \bibinfo {author}
  {\bibfnamefont{K.}~\bibnamefont{Krushelnick}},\ }%
  \bibfield{journal}{%
  \bibinfo {journal} {Phys. Rev. Lett.}\ }%
  \textbf{\bibinfo {volume} {94}},\ \bibinfo {pages} {245001} (\bibinfo {month}
  {Jun}\ \bibinfo {year} {2005})%
  \bibAnnoteFile{NoStop}{Mangles2005}%
\bibitem{Pukhov1999}%
  \BibitemOpen
  \bibfield{author}{%
  \bibinfo {author} {\bibfnamefont{A.}~\bibnamefont{Pukhov}}, \bibinfo {author}
  {\bibfnamefont{Z.-M.}\ \bibnamefont{Sheng}},\ and\ \bibinfo {author}
  {\bibfnamefont{J.}~\bibnamefont{Meyer-ter Vehn}},\ }%
  \bibfield{journal}{%
  \bibinfo {journal} {Physics of Plasmas}\ }%
  \textbf{\bibinfo {volume} {6}},\ \bibinfo {pages} {2847} (\bibinfo {year}
  {1999})%
  \bibAnnoteFile{NoStop}{Pukhov1999}%
\bibitem{Arefiev2012}%
  \BibitemOpen
  \bibfield{author}{%
  \bibinfo {author} {\bibfnamefont{A.~V.}\ \bibnamefont{Arefiev}}, \bibinfo
  {author} {\bibfnamefont{B.~N.}\ \bibnamefont{Breizman}}, \bibinfo {author}
  {\bibfnamefont{M.}~\bibnamefont{Schollmeier}},\ and\ \bibinfo {author}
  {\bibfnamefont{V.~N.}\ \bibnamefont{Khudik}},\ }%
  \bibfield{journal}{%
  \bibinfo {journal} {Phys. Rev. Lett.}\ }%
  \textbf{\bibinfo {volume} {108}},\ \bibinfo {pages} {145004} (\bibinfo
  {month} {Apr}\ \bibinfo {year} {2012})%
  \bibAnnoteFile{NoStop}{Arefiev2012}%
\bibitem{Meyer-ter-Vehn1999}%
  \BibitemOpen
  \bibfield{author}{%
  \bibinfo {author} {\bibfnamefont{J.}~\bibnamefont{Meyer-ter Vehn}}\ and\
  \bibinfo {author} {\bibfnamefont{Z.~M.}\ \bibnamefont{Sheng}},\ }%
  \bibfield{journal}{%
  \bibinfo {journal} {Physics of Plasmas}\ }%
  \textbf{\bibinfo {volume} {6}},\ \bibinfo {pages} {641} (\bibinfo {year}
  {1999})%
  \bibAnnoteFile{NoStop}{Meyer-ter-Vehn1999}%
\bibitem{Sheng2002}%
  \BibitemOpen
  \bibfield{author}{%
  \bibinfo {author} {\bibfnamefont{Z.-M.}\ \bibnamefont{Sheng}}, \bibinfo
  {author} {\bibfnamefont{K.}~\bibnamefont{Mima}}, \bibinfo {author}
  {\bibfnamefont{Y.}~\bibnamefont{Sentoku}}, \bibinfo {author}
  {\bibfnamefont{M.~S.}\ \bibnamefont{Jovanovic}}, \bibinfo {author}
  {\bibfnamefont{T.}~\bibnamefont{Taguchi}}, \bibinfo {author}
  {\bibfnamefont{J.}~\bibnamefont{Zhang}},\ and\ \bibinfo {author}
  {\bibfnamefont{J.~M.}\ \bibnamefont{ter Vehn}},\ }%
  \bibfield{journal}{%
  \bibinfo {journal} {Phys.Rev.Lett.}\ }%
  \textbf{\bibinfo {volume} {88}},\ \bibinfo {pages} {055004} (\bibinfo {year}
  {2002})%
  \bibAnnoteFile{NoStop}{Sheng2002}%
\bibitem{Chen2006}%
  \BibitemOpen
  \bibfield{author}{%
  \bibinfo {author} {\bibfnamefont{M.}~\bibnamefont{Chen}}, \bibinfo {author}
  {\bibfnamefont{Z.-M.}\ \bibnamefont{Sheng}}, \bibinfo {author}
  {\bibfnamefont{J.}~\bibnamefont{Zheng}}, \bibinfo {author}
  {\bibfnamefont{Y.-Y.}\ \bibnamefont{Ma}}, \bibinfo {author}
  {\bibfnamefont{M.}~\bibnamefont{Bari}}, \bibinfo {author}
  {\bibfnamefont{Y.-T.}\ \bibnamefont{Li}},\ and\ \bibinfo {author}
  {\bibfnamefont{J.}~\bibnamefont{Zhang}},\ }%
  \bibfield{journal}{%
  \bibinfo {journal} {Opt. Express}\ }%
  \textbf{\bibinfo {volume} {14}},\ \bibinfo {pages} {3093} (\bibinfo {month}
  {Apr}\ \bibinfo {year} {2006})%
  \bibAnnoteFile{NoStop}{Chen2006}%
\bibitem{Gallard2011}%
  \BibitemOpen
  \bibfield{author}{%
  \bibinfo {author} {\bibfnamefont{S.~A.}\ \bibnamefont{Gaillard}}, \bibinfo
  {author} {\bibfnamefont{T.}~\bibnamefont{Kluge}}, \bibinfo {author}
  {\bibfnamefont{K.~A.}\ \bibnamefont{Flippo}}, \bibinfo {author}
  {\bibfnamefont{M.}~\bibnamefont{Bussmann}}, \bibinfo {author}
  {\bibfnamefont{B.}~\bibnamefont{Gall}}, \bibinfo {author}
  {\bibfnamefont{T.}~\bibnamefont{Lockard}}, \bibinfo {author}
  {\bibfnamefont{M.}~\bibnamefont{Geissel}}, \bibinfo {author}
  {\bibfnamefont{D.~T.}\ \bibnamefont{Offermann}}, \bibinfo {author}
  {\bibfnamefont{M.}~\bibnamefont{Schollmeier}}, \bibinfo {author}
  {\bibfnamefont{Y.}~\bibnamefont{Sentoku}},\ and\ \bibinfo {author}
  {\bibfnamefont{T.~E.}\ \bibnamefont{Cowan}},\ }%
  \bibfield{journal}{%
  \bibinfo {journal} {Physics of Plasmas}\ }%
  \textbf{\bibinfo {volume} {18}},\ \bibinfo {pages} {056710} (\bibinfo {year}
  {2011})%
  \bibAnnoteFile{NoStop}{Gallard2011}%
\bibitem{Paradkar2012}%
  \BibitemOpen
  \bibfield{author}{%
  \bibinfo {author} {\bibfnamefont{B.~S.}\ \bibnamefont{Paradkar}}, \bibinfo
  {author} {\bibfnamefont{S.~I.}\ \bibnamefont{Krasheninnikov}},\ and\ \bibinfo
  {author} {\bibfnamefont{F.~N.}\ \bibnamefont{Beg}},\ }%
  \bibfield{journal}{%
  \bibinfo {journal} {Physics of Plasmas}\ }%
  \textbf{\bibinfo {volume} {19}},\ \bibinfo {pages} {060703} (\bibinfo {year}
  {2012})%
  \bibAnnoteFile{NoStop}{Paradkar2012}%
\bibitem{Naseri2012}%
  \BibitemOpen
  \bibfield{author}{%
  \bibinfo {author} {\bibfnamefont{N.}~\bibnamefont{Naseri}}, \bibinfo {author}
  {\bibfnamefont{D.}~\bibnamefont{Pesme}}, \bibinfo {author}
  {\bibfnamefont{W.}~\bibnamefont{Rozmus}},\ and\ \bibinfo {author}
  {\bibfnamefont{K.}~\bibnamefont{Popov}},\ }%
  \bibfield{journal}{%
  \bibinfo {journal} {Phys. Rev. Lett.}\ }%
  \textbf{\bibinfo {volume} {108}},\ \bibinfo {pages} {105001} (\bibinfo
  {month} {Mar}\ \bibinfo {year} {2012})%
  \bibAnnoteFile{NoStop}{Naseri2012}%
\bibitem{Krygier2014}%
  \BibitemOpen
  \bibfield{author}{%
  \bibinfo {author} {\bibfnamefont{A.~G.}\ \bibnamefont{Krygier}}, \bibinfo
  {author} {\bibfnamefont{D.~W.}\ \bibnamefont{Schumacher}},\ and\ \bibinfo
  {author} {\bibfnamefont{R.~R.}\ \bibnamefont{Freeman}},\ }%
  \bibfield{journal}{%
  \bibinfo {journal} {Physics of Plasmas (1994-present)}\ }%
  \textbf{\bibinfo {volume} {21}},\ \bibinfo {pages} {023112} (\bibinfo {year}
  {2014})%
  \bibAnnoteFile{NoStop}{Krygier2014}%
\bibitem{Robinson2013}%
  \BibitemOpen
  \bibfield{author}{%
  \bibinfo {author} {\bibfnamefont{A.~P.~L.}\ \bibnamefont{Robinson}}, \bibinfo
  {author} {\bibfnamefont{A.~V.}\ \bibnamefont{Arefiev}},\ and\ \bibinfo
  {author} {\bibfnamefont{D.}~\bibnamefont{Neely}},\ }%
  \bibfield{journal}{%
  \bibinfo {journal} {Phys. Rev. Lett.}\ }%
  \textbf{\bibinfo {volume} {111}},\ \bibinfo {pages} {065002} (\bibinfo
  {month} {Aug}\ \bibinfo {year} {2013})%
  \bibAnnoteFile{NoStop}{Robinson2013}%
\bibitem{Arefiev2014}%
  \BibitemOpen
  \bibfield{author}{%
  \bibinfo {author} {\bibfnamefont{A.~V.}\ \bibnamefont{Arefiev}}, \bibinfo
  {author} {\bibfnamefont{V.~N.}\ \bibnamefont{Khudik}},\ and\ \bibinfo
  {author} {\bibfnamefont{M.}~\bibnamefont{Schollmeier}},\ }%
  \bibfield{journal}{%
  \bibinfo {journal} {Physics of Plasmas}\ }%
  \textbf{\bibinfo {volume} {21}},\ \bibinfo {pages} {033104} (\bibinfo {year}
  {2014})%
  \bibAnnoteFile{NoStop}{Arefiev2014}%
\bibitem{Arefiev2015}%
  \BibitemOpen
  \bibfield{author}{%
  \bibinfo {author} {\bibfnamefont{A.~V.}\ \bibnamefont{Arefiev}}, \bibinfo
  {author} {\bibfnamefont{G.~E.}\ \bibnamefont{Cochran}}, \bibinfo {author}
  {\bibfnamefont{D.~W.}\ \bibnamefont{Schumacher}}, \bibinfo {author}
  {\bibfnamefont{A.~P.~L.}\ \bibnamefont{Robinson}},\ and\ \bibinfo {author}
  {\bibfnamefont{G.}~\bibnamefont{Chen}},\ }%
  \bibfield{journal}{%
  \bibinfo {journal} {Physics of Plasmas}\ }%
  \textbf{\bibinfo {volume} {22}},\ \bibinfo {pages} {013103} (\bibinfo {year}
  {2015})%
  \bibAnnoteFile{NoStop}{Arefiev2015}%
\bibitem{Boyd2003}%
  \BibitemOpen
  \bibfield{author}{%
  \bibinfo {author} {\bibfnamefont{T.~J.~M.}\ \bibnamefont{Boyd}}\ and\
  \bibinfo {author} {\bibfnamefont{J.~J.}\ \bibnamefont{Sanderson}},\ }%
  \emph{\bibinfo {title} {The Physics of Plasmas}}\ (\bibinfo {publisher}
  {Cambridge University Press},\ \bibinfo {year} {2003})%
  \bibAnnoteFile{NoStop}{Boyd2003}%
\bibitem{Pukhov2003}%
  \BibitemOpen
  \bibfield{author}{%
  \bibinfo {author} {\bibfnamefont{A.}~\bibnamefont{Pukhov}},\ }%
  \bibfield{journal}{%
  \bibinfo {journal} {Rep. Prog. Phys.}\ }%
  \textbf{\bibinfo {volume} {66}},\ \bibinfo {pages} {47} (\bibinfo {year}
  {2003})%
  \bibAnnoteFile{NoStop}{Pukhov2003}%
\bibitem{Tsakiris2000}%
  \BibitemOpen
  \bibfield{author}{%
  \bibinfo {author} {\bibfnamefont{G.~D.}\ \bibnamefont{Tsakiris}}, \bibinfo
  {author} {\bibfnamefont{C.}~\bibnamefont{Gahn}},\ and\ \bibinfo {author}
  {\bibfnamefont{V.~K.}\ \bibnamefont{Tripathi}},\ }%
  \bibfield{journal}{%
  \bibinfo {journal} {Physics of Plasmas}\ }%
  \textbf{\bibinfo {volume} {7}},\ \bibinfo {pages} {3017} (\bibinfo {year}
  {2000})%
  \bibAnnoteFile{NoStop}{Tsakiris2000}%
\bibitem{Landau1960}%
  \BibitemOpen
  \bibfield{author}{%
  \bibinfo {author} {\bibfnamefont{L.~D.}\ \bibnamefont{Landau}}\ and\ \bibinfo
  {author} {\bibfnamefont{E.~M.}\ \bibnamefont{Liftshitz}},\ }%
  \emph{\bibinfo {title} {Mechanics}}\ (\bibinfo {publisher} {Butterworth
  Heineman},\ \bibinfo {year} {1960})%
  \bibAnnoteFile{NoStop}{Landau1960}%
\bibitem{Courtois2005}%
  \BibitemOpen
  \bibfield{author}{%
  \bibinfo {author} {\bibfnamefont{C.}~\bibnamefont{Courtois}}, \bibinfo
  {author} {\bibfnamefont{A.~D.}\ \bibnamefont{Ash}}, \bibinfo {author}
  {\bibfnamefont{D.~M.}\ \bibnamefont{Chambers}}, \bibinfo {author}
  {\bibfnamefont{R.~A.~D.}\ \bibnamefont{Grundy}},\ and\ \bibinfo {author}
  {\bibfnamefont{N.~C.}\ \bibnamefont{Woolsey}},\ }%
  \bibfield{journal}{%
  \bibinfo {journal} {Journal of Applied Physics}\ }%
  \textbf{\bibinfo {volume} {98}},\ \bibinfo {pages} {054913} (\bibinfo {year}
  {2005})%
  \bibAnnoteFile{NoStop}{Courtois2005}%
\bibitem{Daido1986}%
  \BibitemOpen
  \bibfield{author}{%
  \bibinfo {author} {\bibfnamefont{H.}~\bibnamefont{Daido}}, \bibinfo {author}
  {\bibfnamefont{F.}~\bibnamefont{Miki}}, \bibinfo {author}
  {\bibfnamefont{K.}~\bibnamefont{Mima}}, \bibinfo {author}
  {\bibfnamefont{M.}~\bibnamefont{Fujita}}, \bibinfo {author}
  {\bibfnamefont{K.}~\bibnamefont{Sawai}}, \bibinfo {author}
  {\bibfnamefont{H.}~\bibnamefont{Fujita}}, \bibinfo {author}
  {\bibfnamefont{Y.}~\bibnamefont{Kitagawa}}, \bibinfo {author}
  {\bibfnamefont{S.}~\bibnamefont{Nakai}},\ and\ \bibinfo {author}
  {\bibfnamefont{C.}~\bibnamefont{Yamanaka}},\ }%
  \bibfield{journal}{%
  \bibinfo {journal} {Phys. Rev. Lett.}\ }%
  \textbf{\bibinfo {volume} {56}},\ \bibinfo {pages} {846} (\bibinfo {month}
  {Feb}\ \bibinfo {year} {1986})%
  \bibAnnoteFile{NoStop}{Daido1986}%
\bibitem{Fujioka2012}%
  \BibitemOpen
  \bibfield{author}{%
  \bibinfo {author} {\bibfnamefont{S.}~\bibnamefont{Fujioka}}, \bibinfo
  {author} {\bibfnamefont{Z.}~\bibnamefont{Zhang}}, \bibinfo {author}
  {\bibfnamefont{N.}~\bibnamefont{Yamamoto}}, \bibinfo {author}
  {\bibfnamefont{S.}~\bibnamefont{Ohira}}, \bibinfo {author}
  {\bibfnamefont{Y.}~\bibnamefont{Fujii}}, \bibinfo {author}
  {\bibfnamefont{K.}~\bibnamefont{Ishihara}}, \bibinfo {author}
  {\bibfnamefont{T.}~\bibnamefont{Johzaki}}, \bibinfo {author}
  {\bibfnamefont{A.}~\bibnamefont{Sunahara}}, \bibinfo {author}
  {\bibfnamefont{Y.}~\bibnamefont{Arikawa}}, \bibinfo {author}
  {\bibfnamefont{K.}~\bibnamefont{Shigemori}}, \bibinfo {author}
  {\bibfnamefont{Y.}~\bibnamefont{Hironaka}}, \bibinfo {author}
  {\bibfnamefont{Y.}~\bibnamefont{Sakawa}}, \bibinfo {author}
  {\bibfnamefont{Y.}~\bibnamefont{Nakata}}, \bibinfo {author}
  {\bibfnamefont{J.}~\bibnamefont{Kawanaka}}, \bibinfo {author}
  {\bibfnamefont{H.}~\bibnamefont{Nagatomo}}, \bibinfo {author}
  {\bibfnamefont{H.}~\bibnamefont{Shiraga}}, \bibinfo {author}
  {\bibfnamefont{N.}~\bibnamefont{Miyanaga}}, \bibinfo {author}
  {\bibfnamefont{T.}~\bibnamefont{Norimatsu}}, \bibinfo {author}
  {\bibfnamefont{H.}~\bibnamefont{Nishimura}},\ and\ \bibinfo {author}
  {\bibfnamefont{H.}~\bibnamefont{Azechi}},\ }%
  \bibfield{journal}{%
  \bibinfo {journal} {Plasma Physics and Controlled Fusion}\ }%
  \textbf{\bibinfo {volume} {54}},\ \bibinfo {pages} {124042}%
  \bibAnnoteFile{NoStop}{Fujioka2012}%
\bibitem{Fujioka2013}%
  \BibitemOpen
  \bibfield{author}{%
  \bibinfo {author} {\bibfnamefont{S.}~\bibnamefont{Fujioka}}, \bibinfo
  {author} {\bibfnamefont{Z.}~\bibnamefont{Zhang}}, \bibinfo {author}
  {\bibfnamefont{K.}~\bibnamefont{Ishihara}}, \bibinfo {author}
  {\bibfnamefont{K.}~\bibnamefont{Shigemori}}, \bibinfo {author}
  {\bibfnamefont{Y.}~\bibnamefont{Hironaka}}, \bibinfo {author}
  {\bibfnamefont{T.}~\bibnamefont{Johzaki}}, \bibinfo {author}
  {\bibfnamefont{A.}~\bibnamefont{Sunahara}}, \bibinfo {author}
  {\bibfnamefont{N.}~\bibnamefont{Yamamoto}}, \bibinfo {author}
  {\bibfnamefont{H.}~\bibnamefont{Nakashima}}, \bibinfo {author}
  {\bibfnamefont{T.}~\bibnamefont{Watanabe}}, \bibinfo {author}
  {\bibfnamefont{H.}~\bibnamefont{Shiraga}}, \bibinfo {author}
  {\bibfnamefont{H.}~\bibnamefont{Nishimura}},\ and\ \bibinfo {author}
  {\bibfnamefont{H.}~\bibnamefont{Azechi}},\ }%
  \bibfield{journal}{%
  \bibinfo {journal} {Sci.Rep.}\ }%
  \textbf{\bibinfo {volume} {3}},\ \bibinfo {pages} {1170} (\bibinfo {year}
  {2013})%
  \bibAnnoteFile{NoStop}{Fujioka2013}%
\bibitem{Albertazzi2014}%
  \BibitemOpen
  \bibfield{author}{%
  \bibinfo {author} {\bibfnamefont{B.}~\bibnamefont{Albertazzi}}, \bibinfo
  {author} {\bibfnamefont{A.}~\bibnamefont{Ciardi}}, \bibinfo {author}
  {\bibfnamefont{M.}~\bibnamefont{Nakatsutsumi}}, \bibinfo {author}
  {\bibfnamefont{T.}~\bibnamefont{Vinci}}, \bibinfo {author}
  {\bibfnamefont{J.}~\bibnamefont{Béard}}, \bibinfo {author}
  {\bibfnamefont{R.}~\bibnamefont{Bonito}}, \bibinfo {author}
  {\bibfnamefont{J.}~\bibnamefont{Billette}}, \bibinfo {author}
  {\bibfnamefont{M.}~\bibnamefont{Borghesi}}, \bibinfo {author}
  {\bibfnamefont{Z.}~\bibnamefont{Burkley}}, \bibinfo {author}
  {\bibfnamefont{S.~N.}\ \bibnamefont{Chen}}, \bibinfo {author}
  {\bibfnamefont{T.~E.}\ \bibnamefont{Cowan}}, \bibinfo {author}
  {\bibfnamefont{T.}~\bibnamefont{Herrmannsdörfer}}, \bibinfo {author}
  {\bibfnamefont{D.~P.}\ \bibnamefont{Higginson}}, \bibinfo {author}
  {\bibfnamefont{F.}~\bibnamefont{Kroll}}, \bibinfo {author}
  {\bibfnamefont{S.~A.}\ \bibnamefont{Pikuz}}, \bibinfo {author}
  {\bibfnamefont{K.}~\bibnamefont{Naughton}}, \bibinfo {author}
  {\bibfnamefont{L.}~\bibnamefont{Romagnani}}, \bibinfo {author}
  {\bibfnamefont{C.}~\bibnamefont{Riconda}}, \bibinfo {author}
  {\bibfnamefont{G.}~\bibnamefont{Revet}}, \bibinfo {author}
  {\bibfnamefont{R.}~\bibnamefont{Riquier}}, \bibinfo {author}
  {\bibfnamefont{H.-P.}\ \bibnamefont{Schlenvoigt}}, \bibinfo {author}
  {\bibfnamefont{I.~Y.}\ \bibnamefont{Skobelev}}, \bibinfo {author}
  {\bibfnamefont{A.}~\bibnamefont{Faenov}}, \bibinfo {author}
  {\bibfnamefont{A.}~\bibnamefont{Soloviev}}, \bibinfo {author}
  {\bibfnamefont{M.}~\bibnamefont{Huarte-Espinosa}}, \bibinfo {author}
  {\bibfnamefont{A.}~\bibnamefont{Frank}}, \bibinfo {author}
  {\bibfnamefont{O.}~\bibnamefont{Portugall}}, \bibinfo {author}
  {\bibfnamefont{H.}~\bibnamefont{Pépin}},\ and\ \bibinfo {author}
  {\bibfnamefont{J.}~\bibnamefont{Fuchs}}\ }%
  \textbf{\bibinfo {volume} {346}},\ \bibinfo {pages} {325} (\bibinfo {year}
  {2014})%
  \bibAnnoteFile{NoStop}{Albertazzi2014}%
\bibitem{Liu2013}%
  \BibitemOpen
  \bibfield{author}{%
  \bibinfo {author} {\bibfnamefont{B.}~\bibnamefont{Liu}}, \bibinfo {author}
  {\bibfnamefont{H.~Y.}\ \bibnamefont{Wang}}, \bibinfo {author}
  {\bibfnamefont{J.}~\bibnamefont{Liu}}, \bibinfo {author}
  {\bibfnamefont{L.~B.}\ \bibnamefont{Fu}}, \bibinfo {author}
  {\bibfnamefont{Y.~J.}\ \bibnamefont{Xu}}, \bibinfo {author}
  {\bibfnamefont{X.~Q.}\ \bibnamefont{Yan}},\ and\ \bibinfo {author}
  {\bibfnamefont{X.~T.}\ \bibnamefont{He}},\ }%
  \bibfield{journal}{%
  \bibinfo {journal} {Phys. Rev. Lett.}\ }%
  \textbf{\bibinfo {volume} {110}},\ \bibinfo {pages} {045002} (\bibinfo
  {month} {Jan}\ \bibinfo {year} {2013})%
  \bibAnnoteFile{NoStop}{Liu2013}%
\bibitem{Liu2004}%
  \BibitemOpen
  \bibfield{author}{%
  \bibinfo {author} {\bibfnamefont{H.}~\bibnamefont{Liu}}, \bibinfo {author}
  {\bibfnamefont{X.~T.}\ \bibnamefont{He}},\ and\ \bibinfo {author}
  {\bibfnamefont{S.~G.}\ \bibnamefont{Chen}},\ }%
  \bibfield{journal}{%
  \bibinfo {journal} {Phys. Rev. E}\ }%
  \textbf{\bibinfo {volume} {69}},\ \bibinfo {pages} {066409} (\bibinfo {month}
  {Jun}\ \bibinfo {year} {2004})%
  \bibAnnoteFile{NoStop}{Liu2004}%
\end{thebibliography}%

\end{document}